\RequirePackage[T1]{fontenc}
\documentclass[12pt]{article}

\usepackage[height=8.85in,width=6.45in]{geometry}

\usepackage{times,draft}

\usepackage[utf8]{inputenc}
\usepackage{amsmath}
\usepackage{amssymb}
\usepackage{mathtools}
\numberwithin{equation}{section}
\usepackage{slashed}
\usepackage{braket}
\usepackage[svgnames,psnames]{xcolor}
\usepackage{multicol}
\usepackage{multirow}

\usepackage{mdframed}

\usepackage{array}
\usepackage{blkarray}
\usepackage{arydshln}
\usepackage{dsfont}

\newcolumntype{M}[1]{>{\centering\arraybackslash}m{#1}}
\newcolumntype{N}{@{}m{0pt}@{}}

\begin{document}

\begin{titlepage}
 	\hfill  	YITP-SB-2023-27, MIT-CTP/5606
 	\\

\title{Quantization of Axion-Gauge Couplings and Non-Invertible Higher Symmetries}

\author{Yichul Choi${}^{1,2}$, Matthew Forslund${}^1$,  Ho Tat Lam${}^3$, and Shu-Heng Shao${}^1$}

		\address{${}^{1}$C.\ N.\ Yang Institute for Theoretical Physics, Stony Brook University\\
        ${}^{2}$Simons Center for Geometry and Physics, Stony Brook University\\
		${}^{3}$Center for Theoretical Physics, Massachusetts Institute of Technology
		}

\abstract

We derive model-independent quantization conditions on the axion couplings  (sometimes known as the anomaly coefficients) to the Standard Model gauge group $[SU(3)\times SU(2)\times U(1)_Y]/\mathbb{Z}_q$ with $q=1,2,3,6$.  
Using these quantization conditions, we prove that any QCD axion model to the right of the $E/N=8/3$ line on the $|g_{a\gamma\gamma}|$-$m_a$ plot   must necessarily face the axion domain wall problem in a post-inflationary scenario. 
We further demonstrate the higher-group and non-invertible global symmetries in the Standard Model coupled to a single axion. 
These generalized global symmetries lead to universal bounds on the axion string tension and the monopole mass.
If the axion were discovered in the future, our quantization conditions could be used to constrain the global form of the Standard Model gauge group.

\end{titlepage}

\eject

\tableofcontents

\section{Introduction}

Axions have long been a major target in particle phenomenology.  
Originally emerging as an elegant solution to the strong CP problem \cite{Weinberg:1977ma,Peccei:1977hh,Wilczek:1977pj,Peccei:1977ur}, axions have since been identified as one of the most well-motivated dark matter candidates~\cite{Preskill:1982cy,Dine:1982ah,Abbott:1982af,Adams:2022pbo}, appearing in a number of extensions of the Standard Model (SM). 
They have motivated dozens of unique experimental searches bridging a wide variety of disciplines such as collider physics, astrophysics, condensed matter physics, and quantum optics~\cite{Adams:2022pbo}. 
Axion-like particles can provide a natural candidate for the inflaton~\cite{Freese:1990rb,Marsh:2015xka}, and may also play a central role in a variety of important cosmological effects such as the cosmological constant problem~\cite{Marsh:2015xka}.
Axion-like particles also exist ubiquitously in string theory \cite{Dine:1986bg,Choi:1985je,Barr:1985hk,Witten:1984dg,Svrcek:2006yi,Arvanitaki:2009fg}.
See, for example, \cite{Kim:2008hd,Marsh:2015xka,DiLuzio:2020wdo,Agrawal:2022yvu,Blinov:2022tfy,Reece:2023czb} for various reviews.

Symmetry and topology have always been at the center stage in axion physics. 
In recent years, there has been a transformative development in our understanding of  symmetries in theoretical physics, motivated by advancements in high energy physics, condensed matter physics, and mathematical physics. 
Global symmetries have been generalized in several different directions, including the higher-form symmetries, higher-group symmetries, non-invertible symmetries, and more. 
(See \cite{McGreevy:2022oyu,Cordova:2022ruw,Brennan:2023mmt,Schafer-Nameki:2023jdn,Bhardwaj:2023kri,Shao:2023gho} for reviews.) 
In particular, the axion shift global symmetry,  a classical   $U(1)$ global symmetry that suffers from the quantum Adler-Bell-Jackiw (ABJ) anomaly \cite{PhysRev.177.2426,Bell:1969ts}, is recently recognized to be an exact non-invertible global symmetry in the axion-Maxwell theory   \cite{Choi:2022jqy,Cordova:2022ieu,Choi:2022fgx,Yokokura:2022alv} (which build on earlier works in \cite{Choi:2021kmx,Kaidi:2021xfk}). 
This is  a new kind of symmetry that does not have an inverse. 
Higher-group symmetries for axions are also found in \cite{Seiberg:2018ntt,Cordova:2019uob,Hidaka:2020iaz,Hidaka:2020izy,Brennan:2020ehu,Brennan:2023kpw}, as well as non-invertible higher-form symmetries \cite{Choi:2022fgx,Yokokura:2022alv}. 
These new symmetries lead to universal inequalities on the energy scales where various emergent global symmetries are broken in the UV completion of  axion models \cite{Brennan:2020ehu,Choi:2022fgx}. In particular,  for axion-electrodynamics, it was shown in \cite{Choi:2022fgx} that the axion string tension $T$ and the monopole mass $m_\text{monopole}$ are bounded from below by the mass of the lightest electrically charged particle $m_\text{electric}$, i.e., $m_\text{electric}\lesssim \min\left\{m_\text{monopole},\sqrt{T}\right\}$.\footnote{Note that the electrically and magnetically charged particles are not on the same footing, i.e., there is no electromagnetic duality. This is because we assume at low energies the effective theory contains an axion field that couples to the gauge field as $\theta F\wedge F$.}

In this paper, we extend the previous analysis in \cite{Brennan:2020ehu,Choi:2022fgx} to the full SM coupled to a single axion field $\theta(x)$. 
The three coupling constants $K_3,K_2,K_1$ (defined in \eqref{eq:axion_coupling_comp}) between the axion field and the instanton number densities of the $su(3)\times su(2) \times u(1)_Y$ gauge fields are quantized by imposing the periodic identification $\theta(x)\sim \theta(x)+2\pi$. 
The precise quantization conditions depend on the global form of the SM gauge group, which we explain below. 
It is known that a $\mathbb{Z}_6$ center subgroup of $SU(3)\times SU(2)\times U(1)_Y$ acts trivially on all the SM particles. 
Therefore, the SM gauge group need not be a product group, but  can  be one of the following:
\begin{equation}
G_{SM} = [SU(3)\times SU(2) \times U(1)_Y]/\mathbb{Z}_q\,,~~~~q=1,2,3,6\,.
\end{equation}
More physically, the global form of $G_{SM}$ depends on what gauge charges are allowed or introduced in physics beyond the Standard Model (BSM).  
For example, in the $SU(5)$ and many other Grand Unified Theories (GUT), all new particles  carry vanishing $\mathbb{Z}_6$ charges, and  $q=6$.\footnote{ 
Said differently,
the SM gauge group is embedded as a $[SU(3)\times SU(2)\times U(1)_Y]/\mathbb{Z}_6 \subset SU(5)$ subgroup, but not as a product group $SU(3)\times SU(2)\times U(1)_Y$.
} 
In other UV models for the axion (such as the simplest version of the KSVZ model where the heavy fermion is in the fundamental of $SU(3)$ but neutral under $SU(2)\times U(1)_Y$ \cite{Kim:1979if,Shifman:1979if}), the heavy fermions may carry nontrivial $\mathbb{Z}_6$ charges, and the corresponding global form of $G_{SM}$ is different.\footnote{The $q=6$ case is in many ways phenomenologically preferred. In particular, if $q\neq6$, the $\mathbb{Z}_6$-charged BSM particles  cannot decay into the SM particles and form stable relics in cosmology. See Section \ref{sec:g} for more discussions. 
For the same reason,  we assume that the hypercharge is never smaller than $1/6$  and the gauge group $U(1)_Y$ is compact. The $q=6$ case can also be realized from F-theory \cite{Cvetic:2017epq,Harlow:2018tng,Raghuram:2019efb}. } 
See \cite{Tong:2017oea} for more discussions.

We provide several  bottom-up derivations of  the quantization conditions of the axion-gauge couplings $K_i$'s for each version of the global form of the SM gauge group.  
Below the electroweak symmetry breaking (EWSB) scale, this gives the quantization conditions of the axion-gluon coupling $N$ and (bare) axion-photon coupling $E$ (defined in \eqref{eq:ew_comp}).\footnote{The couplings $K_i$ (and also $E,N$)  are sometimes referred to as the ``anomaly coefficients'' in the literature. This is accurate if the axion model arises from the Peccei-Quinn (PQ) mechanism where there is a classical global symmetry suffering from an ABJ anomaly. However, there are more general UV realizations of the axion models (such as from string theory or extra dimensions), where   the $K_i$'s might not admit an interpretation in terms of any anomaly coefficient. Since our discussion is universal and does not refer to a particular UV completion of the axion model,  we prefer not to refer to the $K_i$'s as the ``anomaly coefficients.''}   
These quantization conditions are satisfied by every UV consistent axion models, but our derivation is universal and model-independent. We summarize our quantization conditions in Table \ref{table:quantization}.

Our quantization conditions give interesting constraints on the effective axion-photon coupling $g_{a\gamma\gamma}$, which is expressed in terms of the ratio $E/N$. 
We show that, in the $q=6$ case, any QCD axion model lying to the right of the $E/N=8/3$ line (realized by one of the DFSZ models \cite{Zhitnitsky:1980tq,Dine:1981rt}  and many GUT models) on the $|g_{a\gamma\gamma}|$-$m_a$ plot must necessarily have more than one axion domain wall in a post-inflationary scenario. 
Our result provides an  invariant meaning for the ratio $E/N=8/3$, which is  taken as a primary target for experimental detection.

We further  analyze the generalized global symmetries,  including the higher-group and non-invertible symmetries. 
We summarize the generalized global symmetries,  which depend on the number theoretic properties of the quantized axion-gauge couplings $K_i$,  in Table \ref{table:symmetries}.  
In many cases, we find that  one symmetry is subordinate to the other, meaning that the former cannot exist without the latter. 
This hierarchy constrains the energy scales where these symmetries are broken, from which we derive universal inequalities on the axion string tension and the hypercharge monopole mass, generalizing \cite{Brennan:2020ehu,Choi:2022fgx}.

The rest of the paper is organized as follows. 
In Section \ref{sec:quantization}, we provide several different derivations for the model-independent quantization on the couplings between the axion and the SM gauge fields. 
In Section \ref{sec:g}, we apply the quantization conditions to constrain the effective axion-photon coupling $g_{a\gamma\gamma}$. 
Section \ref{sec:symmetry} discusses the generalized global symmetries of the SM coupled a single axion field, and how they can be used to bound the axion string tension and the monopole mass. In Appendix \ref{app:symm}, we provide more details on the higher-group and non-invertible symmetries in the SM coupled to an axion. 
Finally, appendix \ref{app:fractionalK} discusses fractional axion-gauge couplings in the presence of additional topological degrees of freedom.

\section{Quantization of the axion couplings to gauge fields}\label{sec:quantization}

The axion field, denoted as $a$, is a periodic scalar field whose periodicity is given by $a \sim a + 2\pi f$, where $f$ is the axion decay constant.
We consider a generic coupling of the axion field to the SM gauge fields,
\begin{equation} \label{eq:axion_coupling_comp}
    \frac{K_3 g_3^2}{32\pi^2 } {a\over f} G_{\mu\nu}^a \widetilde{G}^{a,\mu\nu} + \frac{K_2 g_2^2}{32\pi^2 } {a\over f} F_{\mu\nu}^a \widetilde{F}^{a,\mu\nu} +  \frac{K_1 g_1^2}{16\pi^2 } \frac{1}{36} {a\over f}  B_{\mu\nu} \widetilde{B}^{\mu\nu} \,.
\end{equation}
Here, $g_i$'s are the gauge coupling constants, and
$G_{\mu\nu}^a$, $F_{\mu\nu}^a$, and $B_{\mu\nu}$ are the $SU(3)$, $SU(2)$ and $U(1)_Y$ field strengths, respectively.
The coupling constants $K_i$'s can take only a quantized set of values, and the precise quantization conditions will be determined below.
The dual field strengths are defined as $\widetilde{G}^a_{\mu\nu} = \frac{1}{2} \epsilon_{\mu\nu\rho\sigma} G^{a,\rho\sigma}$ and similarly for the $SU(2)$ and $U(1)_Y$ field strengths.
The $U(1)_Y$ hypercharge gauge field is normalized such that 
$Q_Y\in \mathbb{Z}/6$.\footnote{For instance, the covariant derivative acting on the left-handed quark doublet is $D_\mu = \partial_\mu - ig_1 \frac{1}{6} B_\mu + \cdots$ where $B_{\mu\nu} = \partial_\mu B_\nu - \partial_\nu B_\mu$.}
Without loss of generality, we assume $K_3\ge0$, which can always be achieved by a field redefinition $a\to -a$.

Below the EWSB scale, \eqref{eq:axion_coupling_comp} reduces to
\begin{equation} \label{eq:ew_comp}
    \frac{N g_3^2}{16\pi^2 } {a\over f} G_{\mu\nu}^a \widetilde{G}^{a,\mu\nu} +  \frac{E e^2}{16\pi^2 } {a\over f} F_{\text{EM},\mu\nu} \widetilde{F}^{\mu\nu}_\text{EM} + \cdots \,,
\end{equation}
 where $\cdots$ denotes the coupling between the axion and the massive gauge fields. 
The electric coupling $e=g_1\cos\theta_W=g_2\sin\theta_W$ is fixed by the coupling $g_1,g_2$ and the weak mixing angle $\theta_W \equiv \text{tan}^{-1}(g_1/g_2)$. 
Here we follow the standard convention in the literature \cite{Srednicki:1985xd} to denote the axion-gluon coupling by   $N$.\footnote{In the literature, the term ``axion-like particle'' (ALP) is sometimes used instead of ``axion'' if $N=0$. We will not make any such distinction.} It is related to $K_3$ (which equals the number $N_\text{DW}$ of axion domain walls) as\footnote{It is somewhat awkward that  $N$ can sometimes be a half integer, while $K_3=N_\text{DW}$ is always a positive integer as we will discuss later. In contrast, the bare axion-photon coupling $E$ is generally not an integer.}
\begin{equation}
K_3= N_\text{DW}=  2N \,.
\end{equation}
The (bare) axion-photon coupling $E$  is related to $K_1,K_2$ as
\begin{equation} \label{eq:E}
   E = \frac{1}{36}(K_1 + 18K_2) \,.
\end{equation}
This can be obtained from the standard relations, $A_{\text{EM},\mu} = \text{sin}\,\theta_W A_\mu^3 + \text{cos}\,\theta_W B_\mu$, $Z_\mu = \text{cos}\,\theta_W A_\mu^3 - \text{sin}\,\theta_W B_\mu$, and $W_\mu^{\pm} = \frac{1}{\sqrt{2}}(A_\mu^1 \mp iA_\mu^2)$.
Here $A_\mu^a$ and $B_\mu$ are the $SU(2)$ and $U(1)_Y$ gauge fields, respectively.

For our purposes, it is convenient to absorb the gauge couplings as well as the axion decay constant by rescaling the fields as
\begin{equation} \label{eq:normalization}
    \theta \equiv \frac{a}{f} \,, \quad F_{3,\mu\nu}^a \equiv g_3 G_{\mu\nu}^a \,, \quad
    F_{2,\mu\nu}^a \equiv g_2 F^a_{\mu\nu}  \,, \quad F_{1,\mu\nu} \equiv \frac{1}{6} g_1 B_{\mu\nu} \,, \quad 
    F_{\mu\nu} \equiv e F_{\text{EM},\mu\nu}   \,,
\end{equation}
where the fields on the right-hand side are normalized to have the canonical kinetic term such as   $-\frac 14 F_\text{EM}^{\mu\nu}F_{\text{EM},\mu\nu}$. 
By an abuse of terminology, we refer to both $\theta$ and $a$ as an axion.
$\theta$ has the periodicity of $2\pi$, $\theta \sim \theta + 2\pi$, which should be viewed as a gauge symmetry. 
The $SU(3)$, $SU(2)$, and $U(1)_Y$ gauge fields in this normalization convention are denoted as $A_{3,\mu}^a$, $A_{2,\mu}^a$, and $A_{1,\mu}$, respectively.

In terms of the differently normalized fields \eqref{eq:normalization}, the axion coupling \eqref{eq:axion_coupling_comp} can be equivalently written as
\begin{equation} \label{eq:axion_coupling_diff}
    \frac{K_3}{8\pi^2} \theta \, \text{Tr} \, F_3 \wedge F_3 +
    \frac{K_2}{8\pi^2} \theta \, \text{Tr} \, F_2 \wedge F_2 +
    \frac{K_1}{8\pi^2} \theta \, F_1 \wedge F_1  \,,
\end{equation}
where we have also adopted the differential form notation.
The traces are taken in the fundamental representation, and the Lie algebra generators are normalized such that $\text{Tr}(T^a T^b) = \frac{1}{2}\delta^{ab}$.
Similarly, the coupling after EWSB \eqref{eq:ew_comp} can be written as
\begin{equation} \label{eq:ew_diff}
    \frac{N}{4\pi^2} \theta \, \text{Tr} \, F_3 \wedge F_3 + \frac{E}{8\pi^2} \theta \, F \wedge F \,.
\end{equation}

\begin{table}[t!]
    \begin{center}
      \begin{tabular}{ |M{0.7cm}|M{4.2cm}|M{4.7cm}|M{4.7cm}|N } 
       \hline
       $q$ & $G_{SM}$ & Quantization of $K_i$'s & Quantization of $N$ and $E$ &\\[20pt] 
       \hline\hline
       1 & $SU(3) \times SU(2) \times U(1)_Y$ & $K_3, K_2, K_1 \in \mathbb{Z}$  & $N \in \frac{1}{2}\mathbb{Z}$, $E \in \frac{1}{36}\mathbb{Z}$ &\\[30pt]
       \hline 
       2 & $SU(3) \times U(2)$ & $K_3, K_2 \in \mathbb{Z}$, $K_1 \in 2\mathbb{Z}$, $2K_2 + K_1 \in 4\mathbb{Z}$ & $N \in \frac{1}{2}\mathbb{Z}$, $E \in \frac{1}{9}\mathbb{Z}$ &\\[30pt]
       \hline 
       3 & $U(3) \times SU(2)$ & $K_3, K_2 \in \mathbb{Z}$, $K_1 \in 3\mathbb{Z}$, $6K_3 + K_1 \in 9\mathbb{Z}$ & $N \in \frac{1}{2}\mathbb{Z}$, $E \in \frac{1}{12}\mathbb{Z}$, $4N + 12E \in 3\mathbb{Z}$  &\\[30pt]
       \hline 
       6 & $S(U(3)\times U(2))$ & $K_3, K_2 \in \mathbb{Z}$, $K_1 \in 6\mathbb{Z}$, $24K_3 + 18K_2 + K_1 \in 36\mathbb{Z}$ & $N \in \frac{1}{2}\mathbb{Z}$, $E \in \frac{1}{3}\mathbb{Z}$, $4N + 3E \in 3\mathbb{Z}$ & \\[30pt]
       \hline
      \end{tabular}
\end{center}
\caption{Quantization conditions of the axion couplings to the gauge fields for the possible global forms of the SM gauge group.}
\label{table:quantization}
\end{table}

We summarize the quantization conditions in Table \ref{table:quantization}.\footnote{The values of the individual $K_i$'s can change under field redefinitions of the fermions of the form $\psi \rightarrow e^{i\theta}\psi $ \cite{Fraser:2019ojt}. Nonetheless, our quantization conditions in Table \ref{table:quantization} are invariant under such field redefinitions. This serves as a consistency check of these conditions.}
We present them both in terms of $K_i$'s as well as in terms of $E$ and $N$.
The latter are more natural below the EWSB scale and are commonly used in the literature.
They are also directly relevant for the axion-photon coupling $g_{a\gamma\gamma}$ discussed in Section \ref{sec:g}.
Interestingly, we find that the quantizations of different $K_i$'s, $E$, and $N$ are sometimes correlated depending on the global form of the SM gauge group.
In the rest of this section, we explain how the conditions are derived.
We provide several independent (but related) derivations of the quantization conditions:
\begin{enumerate}
    \item Assuming  the  PQ mechanism, we derive the quantization conditions from  analyzing the quantum numbers of the PQ fermions.
    \item Fractional instanton numbers. This  extends the analysis in \cite{Anber:2021upc} for $T^4$ to general spacetime manifolds.
    \item  Anomaly vanishing condition of  certain 1-form global symmetries  in  3d Chern-Simons gauge theories.
    \item  Higher-group symmetries.
\end{enumerate}
The first three derivations will be presented in the rest of this section, while 
the last one is discussed in Appendix \ref{app:highergroup}.

In some axion models, such as the DFSZ model \cite{Zhitnitsky:1980tq,Dine:1981rt}, the axion coupling \eqref{eq:ew_comp} below the EWSB scale is a valid description, but at higher energy scales the UV theory is different from the one we consider in \eqref{eq:axion_coupling_comp}.
In such cases, only the couplings $E$ and $N$ are meaningful whereas $K_i$'s are not.
Our results on the quantization conditions of $E$ and $N$ are still valid even in those cases.

We work under the minimal setup where the only degrees of freedom are the SM fields and a single axion field. 
However, in the presence of additional topological degrees of freedom, the axion-gauge couplings $K_i$ can be further fractionalized \cite{Seiberg:2010qd}, which we review in Appendix \ref{app:fractionalK}. Here we assume that there aren't such topological degrees of freedom.

\subsection{Quantum numbers of Peccei-Quinn fermions}

We begin by demonstrating the quantization conditions in Table \ref{table:quantization} by assuming that the couplings between the axion and the SM gauge fields \eqref{eq:axion_coupling_comp} are generated by the spontaneous breaking of a classical $U(1)_{PQ}$ global symmetry in a UV model with additional heavy fermions. 
The axion couplings to the gauge fields are then given by the ABJ triangle anomaly coefficients, from which we derive the quantization conditions. 
Even though this derivation is intuitive, it is not universal and is subject to the above assumption on the UV completion. 
In later subsections, we provide more rigorous, model-independent derivations that reproduce the same results.

We denote an $SU(2)$ irreducible representation by its spin $J=0,\frac 12, 1,\cdots$, and an $SU(3)$ representation by $\{\ell_1,\ell_2\}$ with $\ell_1\ge \ell_2$.  Here  $\ell_i$ is the number of boxes in the $i$-th row of the Young diagram. 
For instance, $\mathbf{3} = \{1,0\}$, $\mathbf{\bar{3}} = \{1,1\}$, and $\mathbf{8}= \{2,1\}$ are the fundamental, anti-fundamental, and adjoint representations, respectively. The hypercharge is normalized so that $Q_Y\in \mathbb{Z}/6$.

Let $\mathcal{X}(\mathbf{R}) \in \mathbb{Z}$ be the $U(1)_{PQ}$ charge of a heavy left-handed Weyl fermion $\mathbf{R}$ whose quantum numbers are $(\ell_i(\mathbf{R}),J(\mathbf{R}),Q_Y(\mathbf{R}))$.
Then, the triangle anomaly coefficients are given by
\begin{align} \label{eq:anom_coeff}
\begin{split}
    K_3 &= 2 \sum_{\mathbf{R}} \mathcal{X}(\mathbf{R}) \, T(\ell_i(\mathbf{R})) \,d(J(\mathbf{R})) \,,\\ 
    K_2 &= 2 \sum_{\mathbf{R}} \mathcal{X}(\mathbf{R})\, T(J(\mathbf{R}))\, d(\ell_i(\mathbf{R})) \,,\\
    K_1 &= \sum_\mathbf{R} \mathcal{X}(\mathbf{R}) \,(6Q_{Y}(\mathbf{R}))^2 \, d(\ell_i(\mathbf{R}))\,d(J(\mathbf{R})) \,.
\end{split}
\end{align}
Here, $T(X)$ and $d(X)$ denote the (Dynkin) index and dimension of a representation $X$, respectively.
If the SM gauge group $G_{SM}$ is given by $[SU(3)\times SU(2)\times U(1)_Y]/\mathbb{Z}_q$, only those representations $\mathbf{R}$ with trivial charge under $\mathbb{Z}_q \subset SU(3)\times SU(2)\times U(1)_Y$ are allowed. See \cite{Baez:2009dj} for related discussions. Depending on the allowed set of representations, $K_i$'s in \eqref{eq:anom_coeff} can take different sets of quantized values.

We first begin with the $q=1$ case, where there is no restriction on the set of  allowed representations.
The index and dimension of arbitrary $SU(2)$ and $SU(3)$ representations are given by
\begin{align} \label{eq:Td}
\begin{split}
    T(J) &= \frac{1}{12}(2J)(2J+1)(2J+2) \,, \quad d(J) = 2J +1 \,, \\
    T(\ell_i) &= \frac{1}{48}(\ell_1 - \ell_2 +1)(\ell_1 +2)(\ell_2+1)(\ell_1^2 + \ell_2^2 + 3\ell_1 - \ell_1 \ell_2) \,,\\
    \quad d(\ell_i) &= \frac{1}{2}(\ell_1 - \ell_2 +1)(\ell_1 +2)(\ell_2+1) \,.
\end{split}
\end{align}
It is straightforward to show that the index is always an integer or a half integer, for instance, by induction.
From \eqref{eq:anom_coeff}, we obtain
\begin{equation} \label{eq:q1_1}
    q=1:~~~K_3, K_2, K_1 \in \mathbb{Z}  \,.
\end{equation}
In terms of the axion couplings below the EWSB scale, this leads to
\begin{equation} \label{eq:q1_2}
 q=1:~~~   N \in \frac{1}{2} \mathbb{Z} \,, \quad E \in \frac{1}{36} \mathbb{Z}  \,.
\end{equation}

Consider now the $q=2$ case, where the SM gauge group becomes $G_{SM} = [SU(3) \times SU(2) \times U(1)_Y]/\mathbb{Z}_2 = SU(3) \times U(2)$.
We denote elements of $SU(3) \times SU(2) \times U(1)_Y$ by a triple $(g,h,e^{i\alpha})$.
Only the representations on which the $\mathbb{Z}_2$ subgroup generated by $(\mathds{1}_{3\times 3}, -\mathds{1}_{2\times 2}, -1)$ acts trivially are allowed.
The element $-\mathds{1}_{2\times 2} \in SU(2)$ is represented by $(-1)^{2J}$ times the identity matrix in the spin $J$ representation, whereas an element $e^{i\alpha} \in U(1)_Y$ acts as $e^{i\alpha (6Q_Y)}$ on a hypercharge $Q_Y$ representation.
Hence, the condition on the allowed representations for $q=2$ is
\begin{equation} \label{eq:q2_rep}
    2J = 6Q_Y \quad \text{mod 2} \,. 
\end{equation}
Combined with \eqref{eq:anom_coeff} and \eqref{eq:Td}, this leads to
\begin{equation} \label{eq:q2_1}
  q=2:~~~  K_3, K_2 \in \mathbb{Z}\,, \quad  K_1 \in 2\mathbb{Z} \,, \quad 2K_2 + K_1 \in 4\mathbb{Z}   \,.
\end{equation}
The conditions on individual $K_i$'s are easy to see. 
On the other hand, the last condition follows from the fact that $2\times 2T(J)+(6Q_Y)^2 d(J) $ is divisible by 4 for every representation satisfying \eqref{eq:q2_rep}.
After EWSB, we have
\begin{equation} \label{eq:q2_2}
   q=2:~~~ N \in \frac{1}{2}\mathbb{Z} \,, \quad E\in \frac{1}{9} \mathbb{Z}   \,.
\end{equation}

If $q=3$, the SM gauge group is $G_{SM} = [SU(3)\times SU(2)\times U(1)_Y]/\mathbb{Z}_3 = U(3) \times SU(2)$, and the allowed representations are those on which the $\mathbb{Z}_3 \subset SU(3)\times SU(2) \times U(1)_Y$ subgroup generated by $(e^{4\pi i/3}\mathds{1}_{3\times 3}, \mathds{1}_{2\times 2}, e^{2\pi i/3})$ acts trivially.
The element $e^{4\pi i/3}\mathds{1}_{3\times 3} \in SU(3)$ is represented by $e^{4\pi i (\ell_1 + \ell_2)/3}$ times the identity matrix on the $\{\ell_1, \ell_2 \}$ representation.
Therefore, the allowed representations satisfy
\begin{equation} \label{eq:q3_rep}
    \ell_1 + \ell_2 = 6Q_Y \quad \text{mod 3}\,.
\end{equation}
It follows that
\begin{equation} \label{eq:q3_1}
 q=3:~~~   K_3, K_2 \in \mathbb{Z}\,, \quad K_1 \in 3\mathbb{Z} \,, \quad 6K_3 + K_1 \in 9\mathbb{Z}  \,.
\end{equation}
The last condition is due to the fact that $6\times 2T(\ell_i)+(6Q_Y)^2 d(\ell_i) $ is divisible by 9 for every representation satisfying \eqref{eq:q3_rep}.\footnote{
One way to check this is as follows. Using \eqref{eq:q3_rep}, we first write $6Q_Y = (\ell_1 + \ell_2) + 3x$ where $x$ is an integer. Then, from \eqref{eq:Td}, we have
\begin{equation}
    6\times 2T(\ell_i)+(6Q_Y)^2 d(\ell_i) = \frac{3}{4} f(\ell_i) + 3x g(\ell_i)  \quad \text{mod 9} \,,
\end{equation}
where
\begin{align}
\begin{split}
    f(\ell_i) &\equiv (\ell_1 - \ell_2 +1)(\ell_1 +2)(\ell_2 +1)(\ell_1^2 + \ell_2^2 + \ell_1 \ell_2 + \ell_1) \,, \\
    g(\ell_i) &\equiv (\ell_1 + \ell_2)(\ell_1 - \ell_2 + 1)(\ell_1 +2)(\ell_2 +1) \,.
\end{split}
\end{align}
Since $6\times 2T(\ell_i)+(6Q_Y)^2 d(\ell_i)$ is an integer,  $f(\ell_i) = 0$ mod 4 for all $\ell_i$. To prove $6\times 2T(\ell_i)+(6Q_Y)^2 d(\ell_i) = 0$ mod 9, we need to prove $f(\ell_i) = g(\ell_i) = 0$ mod 3 for all $\ell_i$.
This can be proved by, for instance, first explicitly checking that $f(\ell_i) =0$ mod 3 for $\ell_2 = 0,1,2$, then noting that $f(\ell_1,\ell_2 +3) = f(\ell_1,\ell_2)$ mod 3. The same argument applies to $g(\ell_i)$.
}
After EWSB, we have
\begin{equation} \label{eq:q3_2}
   q=3:~~~ N\in \frac{1}{2} \mathbb{Z} \,, \quad E \in \frac{1}{12} \mathbb{Z} \,, \quad 4N + 12E \in 3\mathbb{Z}\,.
\end{equation}

Finally, for the $q=6$ case, the gauge group is  $G_{SM} = [SU(3)\times SU(2)\times U(1)_Y]/\mathbb{Z}_6 = S(U(3)\times U(2))$.
The representations must be invariant under the $\mathbb{Z}_6 \subset SU(3) \times SU(2) \times U(1)_Y$ subgroup generated by $(e^{2\pi i/3}\mathds{1}_{3\times 3},-\mathds{1}_{2\times 2}, e^{2\pi i/6})$.
This requires
\begin{equation} \label{eq:q6_rep}
     2J = 6Q_Y \quad \text{mod 2}\quad \text{and} \quad \ell_1 + \ell_2 = 6Q_Y \quad \text{mod 3} \,.
\end{equation}
All the  particles in the SM, and also in the $SU(5)$ and various other GUT models, obey \eqref{eq:q6_rep}.
This leads to
\begin{equation} \label{eq:q6_1}
 q=6:~~~   K_3, K_2 \in \mathbb{Z}\,, \quad K_1 \in 6\mathbb{Z} \,, \quad 24K_3 + 18K_2 + K_1 \in 36\mathbb{Z} \,.
\end{equation}
Below the EWSB scale, we have
\begin{equation} \label{eq:q6_2}
 q=6:~~~   N \in \frac{1}{2} \mathbb{Z} \,, \quad E \in \frac{1}{3} \mathbb{Z} \,, \quad 4N + 3E \in 3 \mathbb{Z}  \,.
\end{equation}
The correlated quantization condition between $N$ and $E$, which comes from $24K_3 + 18K_2 + K_1 \in 36\mathbb{Z}$, will play an important role in putting constraints on the effective axion-photon coupling $g_{a\gamma\gamma}$ discussed in Section \ref{sec:g}.
Proving $24K_3 + 18K_2 + K_1 \in 36\mathbb{Z}$ is equivalent to showing that $24K_3 + 18K_2 + K_1 = 0$ both mod $4$ and mod $9$, since $36 = 4\cdot 9$ and $\text{gcd}(4,9) =1$.
To check these conditions from \eqref{eq:anom_coeff}, it is sufficient to consider the contribution from a single heavy PQ fermion whose quantum numbers $(\ell_i,J,Q_Y)$ are subject to the condition \eqref{eq:q6_rep}. Indeed, it is straightforward to show that
\begin{align}
\begin{split}
24\times 2T(\ell_i)d(J) 
     &+ 18 \times 2T(J) d(\ell_i) +(6Q_Y)^2 d(\ell_i)d(J) 
    \\&=
    \begin{cases}
      ~~  \left[\,
      2 \times 2T(J)+(6Q_Y)^2 d(J)  \,\right] d(\ell_i) = 0\quad \text{mod 4} \,, 
    \\
    ~~ \left[\,
    6 \times 2T(\ell_i)+ (6Q_Y)^2 d(\ell_i) \, \right] d(J) = 0\quad \text{mod 9} \,.
    \end{cases}
\end{split}
\end{align}
The first condition follows from $2J = 6Q_Y$ mod 2, and the second from $\ell_1 + \ell_2 = 6Q_Y$ mod 3.
Note that \eqref{eq:q6_1} is also equivalent to imposing the $q=2$ and $q=3$ conditions \eqref{eq:q2_1}, \eqref{eq:q3_1} simultaneously.

\subsection{Fractional instantons}\label{sec:fractional}

We now provide  a rigorous, model-independent derivation of the quantization conditions from the fractional instanton numbers.
This derivation is universal and does not require any assumption about the UV origin of the axion. 
The fractional instantons on $T^4$ of the SM gauge group have been previously studied in \cite{Anber:2021upc}   by imposing the 't Hooft's twisted boundary conditions \cite{tHooft:1979rtg,vanBaal:1982ag}.

The periodic identification of the dynamical axion field $\theta\sim \theta+2\pi$ is a gauge symmetry, and the exponentiated action  $e^{iS}$ should be gauge-invariant. In other words, $e^{iS}$ should be single-valued under $\theta\sim \theta+2\pi$.  
In the presence of the axion coupling 
\eqref{eq:axion_coupling_diff}, $e^{iS}$ transforms under $\theta \sim \theta+2\pi$ by a phase $\exp(2\pi i \sum\limits_i K_i n_i)$, where $n_i$'s are the instanton numbers (see \eqref{eq:instantons} below). 
The quantization of the instanton numbers (which depends on the global form of the SM gauge group) then gives the quantization of the $K_i$'s, and therefore  $E,N$.

We begin with $q=1$.
The $SU(3)$, $SU(2)$, and $U(1)_Y$ instanton numbers in this case are not correlated, and they separately satisfy\footnote{Throughout the paper, we assume that the spacetime manifold is spin, which is natural due to the presence of fermions. On non-spin manifolds, we have $n_1 \in \frac{1}{2} \mathbb{Z}$. It is worth mentioning that, however, the existence of fermions in the theory does not necessarily require the spacetime manifold to be spin, since the theory may admit more involved spacetime tangential structures such as spin$^c$. We do not explore these more general possibilities for the SM in this work. See \cite{Wang:2021vki,Wang:2021ayd} for related discussions in the context of the SM and GUT models.
}
\begin{equation} \label{eq:instantons}
    n_3 \equiv \frac{1}{8\pi^2} \int \text{Tr} \, F_3 \wedge F_3 \in \mathbb{Z} \,, \quad
    n_2 \equiv \frac{1}{8\pi^2} \int \text{Tr} \, F_2 \wedge F_2 \in \mathbb{Z} \,, \quad
    n_1 \equiv \frac{1}{8\pi^2} \int F_1 \wedge F_1 \in \mathbb{Z} \,.
\end{equation}
Therefore, $e^{iS}$ is single-valued under the identification $\theta \sim \theta + 2\pi$ for all possible instanton configurations if and only if the quantization condition \eqref{eq:q1_1} is satisfied.

For $q=2$, since the SM gauge group is $G_{SM} = [SU(3) \times SU(2) \times U(1)_Y]/\mathbb{Z}_2 = SU(3) \times U(2)$, the two gauge fields $A_2$ and $A_1$ combine into a single $U(2)$ gauge field given by $A_{U(2)} \equiv A_2 + A_1 \mathds{1}_{2\times 2}$.
It follows that the instanton numbers $n_2$ and $n_1$ are generally fractional, with the fractional parts   correlated as follows:
\begin{equation} \label{eq:q2_inst1}
    n_2 \in \frac{1}{2} \mathbb{Z} \,, \quad n_1 \in \frac{1}{4} \mathbb{Z} \,, \quad n_2 - 2n_1 \in \mathbb{Z} \,.
\end{equation}
To derive \eqref{eq:q2_inst1}, recall that the first and second Chern classes for the $U(2)$ gauge field $A_{U(2)}$ are 
\begin{align}
\begin{split}
  c_1(F_{U(2)}) &= \frac{1}{2\pi} \text{Tr} \, F_{U(2)} = \frac{1}{2\pi} (2F_1) \,, \\
  c_2(F_{U(2)}) &= \frac{1}{8\pi^2} \left[
    \text{Tr} \, F_{U(2)}^2 - \left( \text{Tr} \, F_{U(2)} \right)^2
  \right] = \frac{1}{8\pi^2} \text{Tr} \, F_2^2 - \frac{2}{8\pi^2} F_1^2 \,,
\end{split}
\end{align}
where $F_{U(2)}$ is the $U(2)$ field strength.
The Chern classes satisfy\footnote{The condition $\frac{1}{2} \int c_1 (F_{U(2)}) \wedge c_1 (F_{U(2)}) \in \mathbb{Z}$ is valid only on spin manifolds. On non-spin manifolds, we instead have $\int c_1 (F_{U(2)}) \wedge c_1 (F_{U(2)}) \in \mathbb{Z}$.}
\begin{equation} \label{eq:chern}
    \int c_2 (F_{U(2)}) \in \mathbb{Z} \,, \quad \frac{1}{2} \int c_1 (F_{U(2)}) \wedge c_1 (F_{U(2)}) \in \mathbb{Z} \,,
\end{equation}
from which \eqref{eq:q2_inst1} follows straightforwardly.
By rewriting 
$K_2 n_2 + K_1 n_1$ as $K_2 (n_2 - 2n_1) +  \frac{K_1 + 2K_2}{4} (4n_1)$, we find that 
$e^{iS}$ is single-valued if and only if the quantization condition \eqref{eq:q2_1} is satisfied.

There is an alternative way to find the allowed values of fractional instanton numbers \eqref{eq:q2_inst1}.
First, we may view $A_2$ and $A_1$ separately as $SO(3) = SU(2)/\mathbb{Z}_2$ and $U(1)/\mathbb{Z}_2$ gauge fields, respectively.
By saying that $A_1$ is a $U(1)/\mathbb{Z}_2$ gauge field, we mean that $2A_1$ is a properly normalized $U(1)$ gauge field satisfying the usual quantization of magnetic fluxes $\oint \frac{2F_1}{2\pi} \in \mathbb{Z}$.
The fact that the two gauge fields combine into the $U(2)$ gauge field  $A_{U(2)} = A_2 + A_1 \mathds{1}_{2\times 2}$ further means that \cite{Cordova:2019uob,Hsin:2020nts} 
\begin{equation} \label{eq:q2_classes}
   c_1 (F_1) =  w_2 (A_2) \quad \text{mod 2} \,.
\end{equation}
Here the first Chern class $c_1 (F_1) = \frac{2F_1}{2\pi}$ is viewed as an element of the cohomology group $H^2(X,\mathbb{Z})$ where $X$ is the spacetime manifold.  $w_2(A_2) \in H^2(X,\mathbb{Z}_2)$ is the second Stiefel-Whitney class for $A_2$, which measures the obstruction to lift an $SO(3)$ bundle to an $SU(2)$ bundle.\footnote{ 
An exposition for the Stiefel-Whitney classes aimed at physicists can be found, for instance, in \cite{Aharony:2013hda,Nakahara:2003nw}.}  
The relation \eqref{eq:q2_classes} guarantees that the transition functions of the $SO(3)$ and $U(1)/\mathbb{Z}_2$ bundles are correlated in such a way that they consistently combine into a $U(2)$ bundle.
Moreover, the fractional part of the instanton number for the $SO(3)$ gauge field $A_2$ can be determined from the second Stiefel-Whitney class as \cite{Aharony:2013hda,Kapustin:2014gua,Gaiotto:2014kfa,Hsin:2018vcg,Cordova:2019uob,Hsin:2020nts}
\begin{equation} \label{eq:q2_frac}
    n_2 = \frac{1}{4} \int \mathcal{P}(w_2(A_2)) \quad \text{mod 1} \,.
\end{equation}
$\mathcal{P}: H^2(X,\mathbb{Z}_2) \rightarrow H^4(X,\mathbb{Z}_{4})$ is the Pontryagin square, which is reviewed, for instance, in \cite[Appendix B]{Hsin:2020nts}.
On spin manifolds, we have $\int \mathcal{P}(w_2(A_2))=0, 2$ mod 4.
From these, we can recover \eqref{eq:q2_inst1} as follows.
First, the fact that $c_1(F_1) = \frac{2F_1}{2\pi}$ has integral periods implies that $n_1 = \frac{1}{8} \int c_1(F_1) \wedge c_1(F_1) \in \frac{1}{4} \mathbb{Z}$.
Next, \eqref{eq:q2_frac} together with the fact that $\int \mathcal{P}(w_2(A_2))=0, 2$ mod 4 gives us $n_2 \in \frac{1}{2}\mathbb{Z}$.
Lastly, taking the Pontryagin square of both sides of \eqref{eq:q2_classes} and integrating them over the spacetime manifold leads to the correlated condition $n_2 - 2n_1 \in \mathbb{Z}$.

We now move on to the $q=3$ case. 
The SM gauge group is $G_{SM} = [SU(3)\times SU(2)\times U(1)_Y]/\mathbb{Z}_3 = U(3) \times SU(2)$, and the two gauge fields $A_3$ and $A_1$ combine into a $U(3)$ gauge field $A_{U(3)} \equiv A_3 + A_1 \mathds{1}_{3\times 3}$.
The instanton numbers $n_3$ and $n_1$ can take fractional values and they satisfy
\begin{equation} \label{eq:q3_inst1}
    n_3 \in \frac{1}{3} \mathbb{Z} \,, \quad n_1 \in \frac{1}{9} \mathbb{Z} \,, \quad n_3 - 6n_1 \in \mathbb{Z}  \,,
\end{equation}
Similar to before, \eqref{eq:q3_inst1} is most easily understood in terms of the Chern classes of the $U(3)$ gauge field.
We have
\begin{align}
\begin{split}
  c_1(F_{U(3)}) &= \frac{1}{2\pi} \text{Tr} \, F_{U(3)} = \frac{1}{2\pi} (3F_1) \,, \\
  c_2(F_{U(3)}) &= \frac{1}{8\pi^2} \left[
    \text{Tr} \, F_{U(3)}^2 - \left( \text{Tr} \, F_{U(3)} \right)^2
  \right] = \frac{1}{8\pi^2} \text{Tr} \, F_3^2 - \frac{6}{8\pi^2} F_1^2 \,.
\end{split}
\end{align}
The quantization \eqref{eq:chern} then leads to  \eqref{eq:q3_inst1}.
By rewriting  $K_3 n_3 + K_1 n_1$ as  $K_3 (n_3- 6n_1) + \frac{K_1 + 6K_3}{9} (9n_1)$, 
we see that $e^{iS}$ is single-valued if and only if \eqref{eq:q3_1} is satisfied.

Alternatively, to derive \eqref{eq:q3_inst1}, we may first view $A_3$ and $A_1$ as $PSU(3) = SU(3)/\mathbb{Z}_3$ and $U(1)/\mathbb{Z}_3$ gauge fields, respectively.  
The condition that they combine into a $U(3)$ gauge field implies that
\begin{equation} \label{eq:q3_classes}
    c_1(F_1) = w_2(A_3) \quad \text{mod 3} \,,
\end{equation}
where the first Chern class $c_1(F_1) = \frac{3F_1}{2\pi}$ for $A_1$ is again viewed as an element of $H^2(X,\mathbb{Z})$. 
Similarly, the fractional part of the $PSU(3)$ instanton number is determined by $w_2(A_3)$ as \cite{Aharony:2013hda,Kapustin:2014gua,Gaiotto:2014kfa,Hsin:2018vcg,Cordova:2019uob,Hsin:2020nts}
\begin{equation} \label{eq:q3_frac}
     n_3 = \frac{1}{3} \int w_2(A_3) \cup w_2(A_3) \quad \text{mod 1} \,.
\end{equation}
From these, the quantization of the individual fractional instanton numbers, $n_3 \in \frac{1}{3} \mathbb{Z}$ and $n_1 \in \frac{1}{9}\mathbb{Z}$, follows immediately.
The correlated quantization condition $n_3 - 6n_1 \in \mathbb{Z}$ is also easily obtained by taking the square on both sides of \eqref{eq:q3_classes} and then integrating them over the spacetime manifold.

Finally, for $q=6$, the allowed fractional values of the instanton numbers and their correlations are:
\begin{align} \label{eq:q6_inst1}
\begin{split}
    &n_3 \in \frac{1}{3} \mathbb{Z} \,, \quad n_2 \in \frac{1}{2} \mathbb{Z} \,, \quad n_1 \in \frac{1}{36} \mathbb{Z} \,,\\
    &n_3 - 24n_1 \in \mathbb{Z} \,, \quad n_2 - 18n_1 \in \mathbb{Z} \,, \quad 2n_3 + n_2 + 6n_1 \in \mathbb{Z} \,.
\end{split}
\end{align}
(The last condition $2n_3 + n_2 + 6n_1 \in \mathbb{Z}$ is redundant and can be obtained from the other conditions, but we keep it for later convenience.) 
One can derive \eqref{eq:q6_inst1} as follows.
Since now the SM gauge group is $[SU(3)\times SU(2)\times U(1)_Y]/\mathbb{Z}_6$, we may view the gauge fields $A_3$, $A_2$, and $A_1$ as $PSU(3)$, $SO(3)$, and $U(1)/\mathbb{Z}_6$ gauge fields, respectively, where the various characteristic classes are related by \cite{Cordova:2019uob,Hsin:2020nts}
\begin{equation} \label{eq:q6_classes}
    c_1 (F_1) = w_2 (A_2) \quad \text{mod 2} \quad \text{and } \quad c_1(F_1) = w_2(A_3) \quad \text{mod 3} \,.
\end{equation}
The conditions in \eqref{eq:q6_classes} imply that the transition functions of $PSU(3)$, $SO(3)$, and $U(1)/\mathbb{Z}_6$ bundles are correlated in such a way that they consistently combine into a $[SU(3)\times SU(2)\times U(1)_Y]/\mathbb{Z}_6$ bundle.
The properly normalized first Chern class which has integral periods is now $c_1(F_1) = \frac{6F_1}{2\pi}$, and the fractional parts of $SO(3)$ and $PSU(3)$ instanton numbers are again given by \eqref{eq:q2_frac} and \eqref{eq:q3_frac}, respectively. 
The quantizations of the individual instanton numbers in \eqref{eq:q6_inst1} follow immediately, and the correlated quantization conditions $n_3 - 24n_1 \in \mathbb{Z}$ and $n_2 - 18n_1 \in \mathbb{Z}$ are obtained by taking (Pontryagin) squares of the two sides of equations in \eqref{eq:q6_classes} and then integrating them over the spacetime manifold, similar to the $q=2$ and $q=3$ cases.
To obtain the quantization condition of $K_i$'s in \eqref{eq:q6_1}, we can rewrite $K_3 n_3 + K_2 n_2 + K_1 n_1$ as $K_3 (n_3 -24n_1) + K_2 (n_2 -18n_1) + \frac{K_1 + 18K_2 + 24K_3}{36} (36n_1)$.
We see that the exponentiated action $e^{iS}$ is single-valued if and only if \eqref{eq:q6_1} is satisfied.

As a side remark, the conditions on the instanton numbers can be used to derive the periodicities and identifications of the $\theta$-angles in the SM (not coupled to an axion):
\begin{align} \label{eq:theta}
\begin{split}
    &q=1:~~~ (\theta_3, \theta_2, \theta_1) \sim (\theta_3 + 2\pi, \theta_2, \theta_1) \sim (\theta_3, \theta_2+ 2\pi, \theta_1) \sim (\theta_3, \theta_2, \theta_1+ 2\pi) \,,\\
    &q=2:~~~ (\theta_3, \theta_2, \theta_1) \sim (\theta_3 + 2\pi, \theta_2, \theta_1) \sim (\theta_3, \theta_2 + 4\pi, \theta_1) \sim (\theta_3, \theta_2, \theta_1 + 8\pi) \\
    &~~~~~~~~~~~~~~~~~~~~~~~~~~~~~~~~ \sim (\theta_3, \theta_2+ 2\pi, \theta_1 + 4\pi) \,,\\
    &q=3:~~~ (\theta_3, \theta_2, \theta_1) \sim (\theta_3 + 6\pi, \theta_2, \theta_1) \sim (\theta_3, \theta_2 + 2\pi, \theta_1) \sim (\theta_3, \theta_2, \theta_1 + 18\pi) \\
    &~~~~~~~~~~~~~~~~~~~~~~~~~~~~~~~~ \sim (\theta_3 +2\pi, \theta_2 , \theta_1 + 6\pi) \,,\\
    &q=6:~~~ (\theta_3, \theta_2, \theta_1) \sim (\theta_3 + 6\pi, \theta_2, \theta_1) \sim (\theta_3, \theta_2 + 4\pi, \theta_1) \sim (\theta_3, \theta_2, \theta_1 + 72\pi) \\
    &~~~~~~~~~~~~~~~~~~~~~~~~~~~~~~~~ \sim (\theta_3 +2\pi, \theta_2 , \theta_1 + 24\pi) \sim (\theta_3 , \theta_2 + 2\pi, \theta_1 + 36\pi) \\
    &~~~~~~~~~~~~~~~~~~~~~~~~~~~~~~~~ \sim (\theta_3 +4\pi, \theta_2 + 2\pi , \theta_1 + 12\pi) \,,
\end{split}
\end{align}
where $\theta_3$, $\theta_2$, and $\theta_1$ are respectively the $SU(3)$, $SU(2)$, and $U(1)_Y$ $\theta$-angles (normalized the same way as in \eqref{eq:axion_coupling_diff}). 
Many of these periodicities  were derived in \cite{Tong:2017oea}. 
Note that depending on the global form of the SM gauge group, $\theta$-angles can have extended as well as correlated periodicities.
Upon EWSB, $\theta$-angle for the electromagnetic gauge field is given by $\theta_{EM} = (\theta_1 + 18\theta_2)/36$, whose (correlated) periodicity can be inferred from \eqref{eq:theta}.
The normalization for $\theta_{EM}$ is such that the Lagrangian contains $\frac{\theta_{EM} e^2}{16\pi^2 } F_{\text{EM},\mu\nu} \widetilde{F}^{\mu\nu}_\text{EM}$ where $F_{\text{EM},\mu\nu}$ is the canonically normalized EM gauge field strength \eqref{eq:normalization}.

\subsection{Chern-Simons gauge theory}

Next, we provide an alternative (but related) derivation of the quantization conditions, which is based on the relationship between the 4d axion couplings and 3d Chern-Simons gauge theories \cite{Witten:1988hf}.
More specifically, we rephrase the quantization condition on $K_i$'s as the vanishing condition of an 't Hooft anomaly of a certain 1-form global symmetry \cite{Gaiotto:2014kfa} in a 3d Chern-Simons theory.
Related discussions can be found, for instance, in \cite{Cordova:2019uob}. 
We emphasize that this derivation is conceptually equivalent to the one in Section \ref{sec:fractional}, but phrased in terms of the anyon spins of a 3d theory.

The main point is that the invariance of the axion couplings to the gauge fields under $\theta\sim \theta+2\pi$ is equivalent to requiring that the  3d Chern-Simons theory 
\begin{equation} \label{eq:CS}
    \mathcal{L}_{CS} = \frac{K_3}{4\pi}\ \text{Tr} \left( A_3 \mathrm{d}A_3 + \frac{2}{3} A_3^3 \right) + \frac{K_2}{4\pi}\ \text{Tr} \left( A_2 \mathrm{d}A_2 + \frac{2}{3} A_2^3 \right) + \frac{K_1}{4\pi} A_1 \mathrm{d}A_1 \,,
\end{equation}
based on the gauge group  $[SU(3)\times SU(2) \times U(1)_Y]/\mathbb{Z}_q$ is well-defined and gauge-invariant.
The Chern-Simons Lagrangian \eqref{eq:CS} is not a globally well-defined differential form, as the gauge fields themselves are not. To define the Chern-Simons action in a mathematically precise way, one extends the Chern-Simons gauge fields to a 4-manifold whose boundary is the spacetime 3-manifold. One then considers a 4d bulk theory with Lagrangian $\mathrm{d}\mathcal{L}_{CS}$, which is a well-defined differential form on the 4-manifold \cite{Dijkgraaf:1989pz}. 
The quantization condition on $K_i$'s follows from requiring that the theory does not depend on the choice of the auxiliary 4-manifold, which in turn is equivalent to requiring $\exp(2\pi i\sum\limits_i K_i n_i) = 1$ on every closed 4-manifolds.
This is analogous to the quantization of the coefficient for the Wess-Zumino-Witten term in the pion Lagrangian \cite{Witten:1983tw}.
Alternatively, the Chern-Simons coupling \eqref{eq:CS} arises from the circle compactification of the axion coupling \eqref{eq:axion_coupling_diff} in the presence of a nontrivial winding number of the axion field around the internal circle.

To find the conditions on $K_i$'s from \eqref{eq:CS}, we begin from $q=1$.
In this case, the Chern-Simons couplings are well-defined for arbitrary integer $K_i$'s.\footnote{On non-spin manifolds, we instead need to require $K_1 \in 2\mathbb{Z}$, whereas $K_2,K_3 \in \mathbb{Z}$ is not affected.}
The theory has a $\mathbb{Z}_3^{(1)} \times \mathbb{Z}_2^{(1)} \times \mathbb{Z}_{|K_1|}^{(1)}$ 1-form center global symmetry.\footnote{In the presence of the Chern-Simons coupling $\frac{K_1}{4\pi} A_1 \mathrm{d}A_1$ in 3d, magnetic monopoles carry electric charges which are multiples of $K_1$, and they break the $U(1)^{(1)}$ center symmetry down to $\mathbb{Z}_{|K_1|}^{(1)}$.}
The $\mathbb{Z}_3^{(1)}$ 1-form symmetry is generated by the $SU(3)$ Wilson line $W_3$ in the representation $\{\ell_1, \ell_2 \} = \{K_3, 0 \}$, the $\mathbb{Z}_2^{(1)}$ 1-form symmetry is generated by the $SU(2)$ Wilson line $W_2$ in the representation $J = K_2/2$, and finally the $\mathbb{Z}_{|K_1|}^{(1)}$ 1-form symmetry is generated by the $U(1)$ Wilson line $W_1 \equiv \exp(i\oint A_1)$.

We  obtain a Chern-Simons theory based on the gauge group $[SU(3)\times SU(2) \times U(1)_Y]/\mathbb{Z}_q$ with $q=2,3,6$ by gauging a $\mathbb{Z}_q^{(1)}$ subgroup of the 1-form symmetry.
For instance, consider the $q=2$ case, where the $A_2$ and $A_1$ gauge fields combine into a $U(2) = [SU(2) \times U(1)_Y]/\mathbb{Z}_2$ gauge field.
Starting from the $q=1$ Chern-Simons theory, we need to gauge the $\mathbb{Z}_2^{(1)} \subset  \mathbb{Z}_3^{(1)} \times \mathbb{Z}_2^{(1)} \times \mathbb{Z}_{|K_1|}^{(1)}$ subgroup generated by the Wilson line $W_2 W_1^{K_1/2}$ to obtain the $q=2$ theory \cite{Seiberg:2016rsg}.
For this to make sense, first of all, we require $K_1 \in 2\mathbb{Z}$, in accordance with \eqref{eq:q2_1}.
Furthermore, to be able to gauge a 1-form symmetry, the corresponding 't Hooft anomaly must vanish.
The 't Hooft anomaly of a $\mathbb{Z}_N^{(1)}$ 1-form symmetry in 3d QFTs is measured by the topological spin of the generating line operator modulo $1/2$ (on spin manifolds) \cite{Hsin:2018vcg}.
In our case, the topological spin of the Wilson line $W_2 W_1^{K_1/2}$ is
\begin{equation}
    \frac{K_2}{4} + \frac{K_1}{8} = \frac{2K_2 + K_1}{8} \quad \text{mod $\frac{1}{2}$} \,.
\end{equation}
Requiring this to vanish, we recover the quantization condition \eqref{eq:q2_1}.

For $q=3$, the two gauge fields $A_3$ and $A_1$ now combine into a $U(3) = [SU(3)\times U(1)_Y]/\mathbb{Z}_3$ gauge field.
Therefore, starting from the $q=1$ Chern-Simons theory, we obtain the $q=3$ theory by gauging the $\mathbb{Z}_3^{(1)} \subset \mathbb{Z}_3^{(1)} \times \mathbb{Z}_2^{(1)} \times \mathbb{Z}_{|K_1|}^{(1)}$ subgroup generated by the Wilson line $W_3 W_1^{K_1/3}$.
This requires $K_1 \in 3\mathbb{Z}$. The 't Hooft anomaly, i.e., the topological spin of the Wilson line $W_3 W_1^{K_1/3}$ is given by
\begin{equation}
    \frac{K_3}{3} + \frac{K_1}{18} = \frac{6K_3 + K_1}{18} \quad \text{mod $\frac{1}{2}$} \,.
\end{equation}
The anomaly vanishes if and only if the quantization condition \eqref{eq:q3_1} is satisfied.

Finally, to obtain the $q=6$ theory, we need to gauge the $\mathbb{Z}_6^{(1)} \subset \mathbb{Z}_3^{(1)} \times \mathbb{Z}_2^{(1)} \times \mathbb{Z}_{|K_1|}^{(1)}$ subgroup generated by the Wilson line $W_3 W_2 W_1^{K_1/6}$, which first requires $K_1 \in 6\mathbb{Z}$.
The topological spin of this Wilson line is
\begin{equation}
    \frac{K_3}{3} + \frac{K_2}{4} + \frac{K_1}{72} = \frac{24K_3 + 18K_2 + K_1}{72} \quad \text{mod $\frac{1}{2}$} \,,
\end{equation}
and it vanishes if and only the quantization condition \eqref{eq:q6_1} is satisfied.

\section{Constraints on the effective axion-photon coupling $g_{a\gamma\gamma}$}\label{sec:g}

We now discuss model-independent constraints on the axion physics from the quantization of the axion couplings to the SM gauge fields. 
We assume that we have a single  QCD axion whose mass $m_a$ comes from the coupling to QCD.
We focus on the effective axion-photon coupling $g_{a\gamma \gamma}$ defined as
\begin{align}
{\cal L}_{a\gamma\gamma}  = - \frac 14 g_{a\gamma\gamma} \,  a F_{\text{EM},\mu\nu}\widetilde F_\text{EM}^{\mu\nu}  \,.
\end{align}
Here $F_{\text{EM},\mu\nu}$ is the field strength for the canonically normalized EM gauge field \eqref{eq:normalization}. 
The effective coupling $g_{a\gamma\gamma}$ receives contribution from the bare coupling $E$ as well as the mixing with $\pi^0$ \cite{GrillidiCortona:2015jxo}:
\begin{align}\label{g}
g_{a\gamma\gamma} = {\alpha\over 2\pi  {f\over 2N}} \left[ {E\over N} -1.92(4)\right]
=  \left[ 
0.203(3) {E\over N} -0.39(1)
\right] {m_a \over \text{GeV}^2}\,,
\end{align}
where $m_a$ is the axion mass and $\alpha=e^2/4\pi$ is the fine structure constant. 
We have written the above expression in the standard notations of $E,N$. 
Recall that $K_3 = N_\text{DW} = 2N$ is the number of axion domain walls.

The quantization condition in Section \ref{sec:quantization} implies that the value of $g_{a\gamma\gamma} /m_a$ cannot be an arbitrary real number, but is subject to some rationality constraint from the ratio $E/N$. 
While any  given real number can be approximated arbitrarily well by a rational number, an accurate approximation    requires a large denominator $N$, which   is related to the number of axion domain walls $N_\text{DW}=2N$.  
In the absence of other mechanisms, stable domain walls  formed after inflation will survive till today and are inconsistent with the current observations. 
Therefore, to avoid the axion domain wall problem in cosmology, the value of $N_\text{DW}$ is preferred to be small in realistic, post-inflationary axion models. 
For any given  upper bound on $N_\text{DW}$, there is a strict lower bound on $|g_{a\gamma\gamma}|/m_a$ from the quantization conditions on $E$ and $N$. 

Another cosmological constraint on the UV completion of the axion model is the stable relic problem~\cite{DiLuzio:2016sbl,DiLuzio:2017pfr}.  
There are typically new   heavy fermions  in UV axion models. 
These heavy particles   have to be able to decay into the SM particles, otherwise they will form stable relics.  
This is  possible if the UV particles carry trivial  gauge charge under the $\mathbb{Z}_6\subset SU(3)\times SU(2)\times U(1)_Y$  subgroup, as all  SM particles do.  
While particles with $\mathbb{Z}_6$ gauge charge can potentially pair annihilate into the SM particles, such scattering processes are suppressed as the universe expands, and some relic density remains.
Therefore, the global form of the SM gauge group is preferred to be $[SU(3)\times SU(2)\times U(1)_Y]/\mathbb{Z}_{q=6}$ in any axion model without the  stable relics in cosmology. 
This is also the gauge group that is compatible with the various GUT models such as $SU(5), Spin(10), E_6$. 
See, for example,  \cite{DiLuzio:2020wdo} and references therein for an extensive discussion on the phenomenological criteria for axion models. 
For this reason, we focus on the $q=6$ case below.

\begin{figure}[t!]
\centering
\includegraphics[width=.7\textwidth]{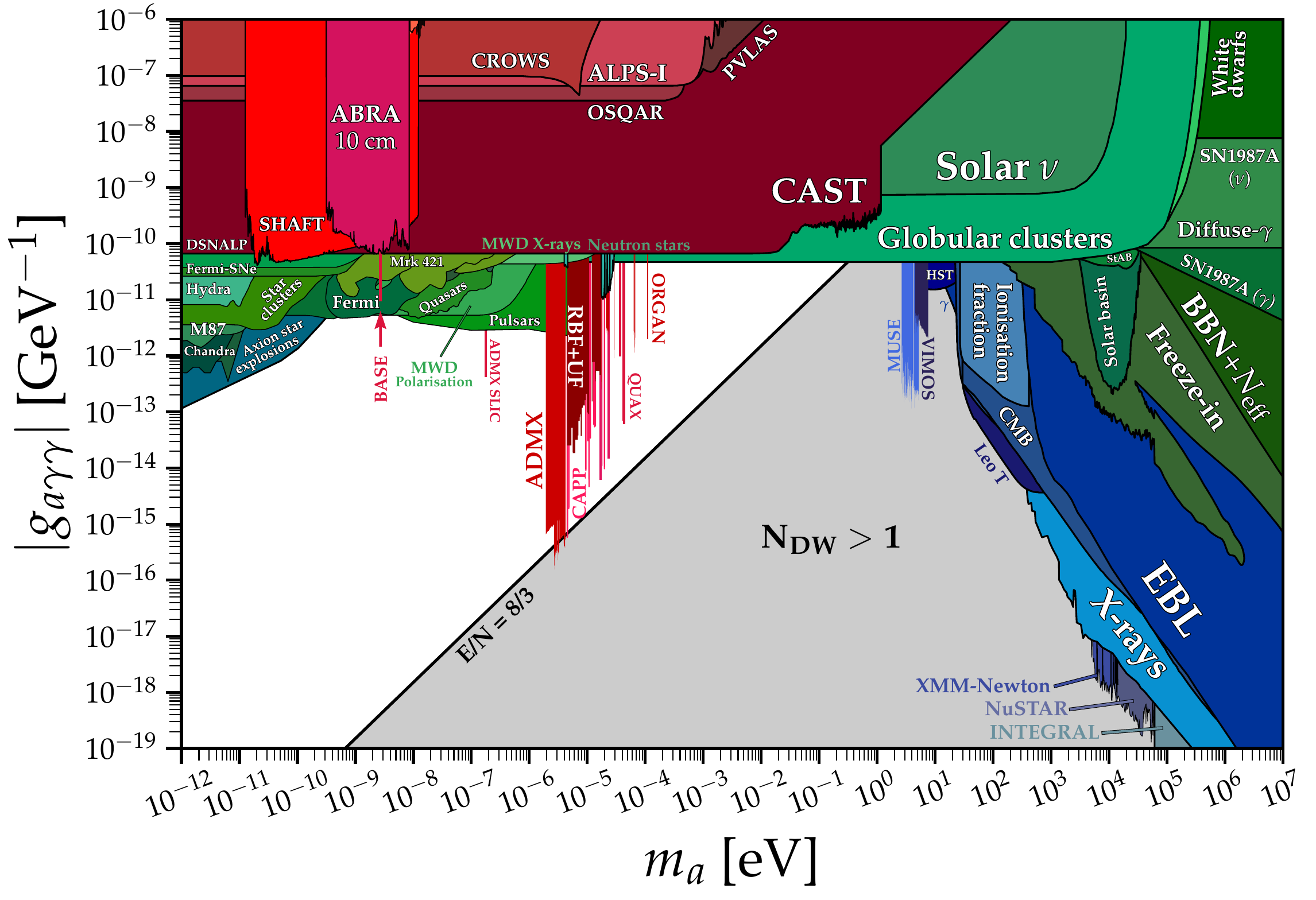}
\caption{Constraints on the effective axion-photon coupling $g_{a\gamma\gamma}$ versus the axion mass $m_a$, modified from~\cite{AxionLimits}. The model-independent quantization conditions \eqref{eq:q6_2} imply that any QCD axion model in the gray region below the $E/N=8/3$ line necessarily faces the axion domain wall problem in a post-inflationary scenario, i.e.,  $N_\text{DW}>1$.}\label{fig:gm}
\end{figure}

The quantization conditions  \eqref{eq:q6_2} for $E$ and $N$ in the $q=6$ case are
 $4N+3E\in 3\mathbb{Z},~~ N\in \frac{1}{2}\mathbb{Z},~~
    E\in \frac{1}{3}\mathbb{Z}$. 
If we want to avoid the domain wall problem, i.e., $N_\text{DW}=2N= 1$, then $g_{a\gamma\gamma}$ in \eqref{g} cannot be arbitrarily small because of the quantization condition \eqref{eq:q6_2}. 
This leads to the following model-independent  lower bound:
\begin{align}\label{bound}
&{ |g_{a\gamma\gamma}|\over m_a } \ge  0.15(1) ~\text{GeV}^{-2}\,,~~~\text{if}~~N_\text{DW}= 1\,.
\end{align}
The lower bound is saturated by 
\begin{equation}
   {E\over N}={8\over3} \,,~~~N_\text{DW}=2N=1,~~~E=\frac 43
\end{equation}
which is the closest rational number $E/N$  to 1.92 subject to the condition \eqref{eq:q6_2} and  $N_\text{DW}=1$. 
This ratio  is famously realized by one of the classic DFSZ models \cite{Zhitnitsky:1980tq,Dine:1981rt}.
(It is worthwhile noting that the standard DFSZ models have $N_\text{DW}=3$ or $6$, while $N_\text{DW}=1$ can be achieved by relaxing PQ charge universality such as in \cite{DiLuzio:2017ogq}.)
It is also the ratio realized by the $SU(5), Spin(10), E_6$ GUT models \cite{Srednicki:1985xd,Agrawal:2022lsp}.
In other words, the region to the right of the $E/N=8/3$ line on the $|g_{a\gamma\gamma}|$-$m_a$ plot must necessarily face the domain wall problem in a post-inflationary scenario (see Figure \ref{fig:gm}, modified from~\cite{AxionLimits}).  
This provides an invariant meaning to the ratio $E/N=8/3$ in the landscape of axion models. 

There are many proposed solutions to the axion domain wall problems in the literature, so models with $N_\text{DW}>1$ are still potentially phenomenologically viable. 
Here we merely point out that those QCD axion models violating the lower bound \eqref{bound}  must necessarily have $N_\text{DW}>1$. 
It should be emphasized that this lower bound relies on the assumption that the axion mass $m_a$ is dominated by the contribution from QCD.\footnote{Going beyond this assumption requires some careful model building to ensure that any additional contributions to the axion potential do not spoil the solution to the strong CP problem. See Section 6.6 of~\cite{DiLuzio:2020wdo} for some examples and discussions.}  
It is also worth pointing out that in the post-inflationary scenario with $N_\text{DW}=1$, there is an upper bound on $f$ to avoid overproducing axions as dark matter from axion string simulations. 
The precise bound is a matter of some debate~\cite{Gorghetto:2020qws,Dine:2020pds,Buschmann:2021sdq}, but conservatively corresponds to $f \lesssim 10^{11}$ GeV.
This would greatly restrict the bottom left region of the $E/N=8/3$ line in Figure~\ref{fig:gm}.\footnote{We thank I.\ Garcia Garcia for discussions on this point.}

As we increase the allowed tolerance of $N_\text{DW}$, eventually the lower bound will be smaller than the uncertainty in \eqref{g}  and cease to be meaningful. 
For instance, the lower bound with $N_\text{DW}\le 2$ is ${ |g_{a\gamma\gamma}|\over m_a } \ge  0.05(1) ~\text{GeV}^{-2}$, which is saturated by $E/N=5/3$.\footnote{Incidentally, the $E/N=5/3$ line appears as the boundary of the ``axion window'' in \cite{DiLuzio:2016sbl,DiLuzio:2017pfr,DiLuzio:2020wdo}.} 
For other values of $q=1,2,3$ (which are not phenomenologically preferred because of  the stable relic problem), the quantization conditions for $E,N$ are different, and the lower bounds on $|g_{a\gamma\gamma}|/m_a$ are saturated by $E/N= 35/18, 2, 13/6$, respectively. However, these lower bounds with $q\neq 6$ are of the same order as the uncertainty in \eqref{g}.  

Finally, it is worth mentioning that in  the simplest version of the KSVZ model \cite{Kim:1979if,Shifman:1979if}, the heavy fermion transforms as $(\mathbf{3},\mathbf{1},0)$ under the $su(3)\times su(2) \times u(1)_Y$ gauge algebra. Therefore, the global form of the SM gauge group must correspond to $q=1$ or $q=2$. To realize the $q=3$ or $q=6$ cases, the heavy fermion must to have a nonzero hypercharge \cite{Reece:2023czb}.
To be more specific, the condition \eqref{eq:q3_rep} requires $6Q_Y = 1$ mod 3 for such a fermion. 
Relatedly, the ratio $E/N = 0$ (which is commonly labeled as the ``KSVZ'' line in the $|g_{a\gamma\gamma}|$-$m_a$ plot) cannot be realized for $q=3$ and $q=6$ by a single heavy fermion transforming under the fundamental of $SU(3)$, because such a fermion must carry a nontrivial electric charge and therefore $E\neq0$.

\section{Higher symmetries of the Standard Model coupled to an axion} \label{sec:symmetry}

In this section we show that the SM coupled to an axion enjoys a myriad of generalized global symmetries, and discuss their implications on the possible UV completions of the theory \cite{Brennan:2020ehu,Choi:2022fgx}.
Many different examples of generalized global symmetries of axion models have been studied in the past, including higher-group symmetries \cite{Seiberg:2018ntt,Cordova:2019uob,Hidaka:2020iaz,Hidaka:2020izy,Brennan:2020ehu}, non-invertible higher-form symmetries \cite{Choi:2022fgx,Yokokura:2022alv}, and non-invertible axion shift symmetries \cite{Choi:2022jqy,Cordova:2022ieu,Choi:2022fgx,Yokokura:2022alv}.

\subsection{Higher-group and non-invertible global symmetries}

We focus on two kinds of generalized global symmetries of the SM coupled to a single axion.
One is the higher-group symmetry which involves the center 1-form symmetry and the winding 2-form symmetry.
We sometimes also refer to the center 1-form symmetry as the electric 1-form symmetry, since it acts on the Wilson lines.
The winding 2-form symmetry comes from the conserved 3-form current $J_{\text{winding}}^{(3)} = \frac{1}{2\pi} \star d\theta$, where the superscript denotes the form degree.
The other generalized symmetry that we consider is the non-invertible 1-form symmetry   discussed in \cite{Choi:2022fgx,Yokokura:2022alv}. 
The non-invertible 1-form symmetry arises from the part of the electric $U(1)^{(1)}$ 1-form symmetry of the free Maxwell theory, which appears to be explicitly broken in the presence of an axion. 
It was realized in \cite{Choi:2022fgx,Yokokura:2022alv} that this $U(1)^{(1)}$ symmetry is not entirely broken by the coupling to the axion; rather, it is resurrected as a non-invertible 1-form symmetry  labeled by $\mathbb{Q}/\mathbb{Z}$ (which is dubbed as the ``non-invertible Gauss law'' in \cite{Choi:2022fgx}). 
The non-invertible 1-form symmetry acts invertibly on the Wilson lines, but non-invertibly on the monopoles and axion strings.

These two generalized symmetries are insensitive to the details of the axion potential and mass, unlike the axion shift symmetry.
Moreover, they are not broken by any higher-dimensional local  operators, which is a general feature of higher-form global symmetries \cite{Gaiotto:2014kfa}.
To break them, one needs to introduce additional dynamical degrees of freedom such as new matter fields that transform nontrivially under the center of the gauge group, dynamical monopoles and/or axion strings.
The existence of these two generalized symmetries is solely determined by the values of $K_i$'s, which is summarized in Table \ref{table:symmetries}.\footnote{Similar to the quantization conditions, the global symmetry structure presented in Table \ref{table:symmetries} is invariant under field redefinitions of fermions.}
We also write the results in terms of the couplings $N$ and $E$ below the EWSB scale.
The derivation of these generalized global symmetries largely follows the formalism in  \cite{Brennan:2020ehu,Choi:2022fgx}, which we review in Appendix \ref{app:symm}.

\begin{table}[t!]
    \begin{center}
      \begin{tabular}{ |M{0.6cm}|M{5.0cm}|M{3.7cm}|M{2.1cm}|M{2.6cm}|N } 
      \hline
       $q$ & \multicolumn{2}{c|}{Higher-Group} &  \multicolumn{2}{c|}{Non-Invertible} &\\[20pt] 
       \hline\hline
       1 & $\dfrac{24K_3 + 18K_2 + K_1}{K}\,$ mod $\,K$ & $\dfrac{48N+36 E}{K}\,$ mod $\,K$ & $K_1$ mod 6& $36E$ mod 6&\\[30pt]
       \hline 
       2 &  $\dfrac{6K_3 + K_1}{(K/2)}\,$ mod $\,\dfrac{K}{2}$ & $\dfrac{12 N + 36 E}{(K/2)}\,$ mod $\,\dfrac{K}{2}$ & $\dfrac{K_1}{2}$ mod 3 & $18E$ mod 3  &\\[30pt]
       \hline 
       3 &  $\dfrac{2K_2 + K_1}{(K/3)}\,$ mod $\,\dfrac{K}{3}$ & $\dfrac{36 E}{(K/3)}\,$ mod $\,\dfrac{K}{3}$ & $\dfrac{K_1}{3}$ mod 2& $12E$ mod 2  &\\[30pt]
       \hline 
       6 & --- &  --- & --- & --- &\\[30pt]
       \hline
      \end{tabular}
\end{center}
\caption{Summary of the higher-group for the center 1-form and winding 2-form symmetries, and non-invertible  1-form symmetries of the SM coupled to an axion, written both in terms of $(K_3,K_2,K_1)$ and $(N,E)$. Whenever an entry in one of the columns is nonzero there exists the corresponding symmetry. Here  $K\equiv \text{gcd}(6,K_1) = \text{gcd}(6,36E)$. Note that $K_1$ and $K$ are always integer multiples of $q$ due to the quantization condition in Table \ref{table:quantization}. The $q=6$ case does not have these generalized global symmetries. }
\label{table:symmetries}
\end{table}

\subsection{Bounds on string tension and monopole mass}

Whenever the values of the axion couplings admit a higher-group or non-invertible symmetry, then there must be a hierarchy to the energy scales at which the different symmetries are broken.
These constraints have been considered for the higher-group symmetries of axion-Yang-Mills, axion-QCD, and axion-Maxwell theory~\cite{Brennan:2020ehu}, as well as for the non-invertible symmetries of axion-Maxwell theory~\cite{Choi:2022fgx}.
In our case the symmetry structure is more complicated but the overall logic remains the same, which we now discuss.

\begin{figure}
\centering
\includegraphics[width=.6\textwidth]{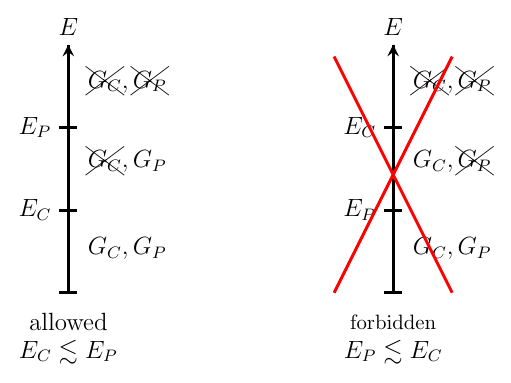}
\caption{For generalized global symmetries, it is common that one ``child'' symmetry $G_C$ cannot exist on its own without a ``parent'' symmetry $G_P$. In axion physics,  $G_C,G_P$ are emergent  in the IR, and this hierarchical structure  constrains the energy scales $E_C,E_P$ these symmetries are broken.}\label{fig:child}
\end{figure}

The generalized global symmetries we discuss are typically emergent in the IR, and they are therefore broken as we go to higher energies. 
However, as a consequence of the higher-group and non-invertible symmetries, they are not all on the same footing. 
Sometimes a ``child'' symmetry $G_C$ is subordinate to a ``parent'' symmetry $G_P$, in the sense that the former symmetry cannot exist without the latter. 
For instance, the child non-invertible symmetry $G_C$ is typically tied to a parent invertible symmetry $G_P$, and the algebra of symmetry elements of $G_C$ contains those of $G_P$.\footnote{The simplest example is the non-invertible symmetry $\mathcal{D}$ in the Ising model (see, for instance, \cite{Petkova:2000ip,Frohlich:2004ef,Aasen:2016dop,Chang:2018iay,Seiberg:2023cdc}). Its algebra $\mathcal{D}\times \mathcal{D}=1+\eta$ involves an invertible parent $\mathbb{Z}_2$ symmetry generated by $\eta$ (i.e., $\eta^2=1$), and therefore the child non-invertible symmetry $\mathcal{D}$ cannot exist without its parent $\eta$.} 
This hierarchical structure constrains the renormalization group flows, since it is not possible to have an effective field theory at an intermediate scale where   the child symmetry is preserved while the parent one is broken. 
This gives universal inequalities  on the energy scales where the symmetries become emergent.  
These inequalities take the form 
$E_C\lesssim E_P$, where $E_C$ and $E_P$ are the energy scales above which the child and parent symmetries are broken, respectively. Since the symmetries are emergent, the inequalities we derive are approximate. 
 See Figure \ref{fig:child}.

Let us first consider the case where there is a nontrivial higher-group symmetry.
There are two invertible symmetries of interest: the $U(1)^{(2)}$ winding 2-form symmetry, and the $\mathbb{Z}_{K/q}^{(1)}$ electric 1-form symmetry, where as usual superscripts denote the form degrees of the global symmetries and $K\equiv \text{gcd}(6,K_1)$.
We denote $E_\text{winding}$ and $E_\text{center}$ as the energy scales below which $U(1)^{(2)}$ and $\mathbb{Z}_{K/q}^{(1)}$ become emergent, respectively.  
Now suppose we flow to a scale below $E_\text{center}$.
Due to the higher-group symmetry structure explained in Appendix \ref{app:symm}, turning on the background gauge field for the $\mathbb{Z}_{K/q}^{(1)}$ electric symmetry automatically activates the background gauge field for the $U(1)^{(2)}$ winding symmetry as well, which only makes sense if we are also below the scale $E_\text{winding}$ \cite{Cordova:2018cvg}. 
In other words, the $\mathbb{Z}_{K/q}^{(1)}$ electric symmetry is the ``child'' symmetry $G_C$ which cannot exist without  the ``parent'' symmetry  $G_P=U(1)^{(2)}$. 
We conclude that
\begin{equation}
    E_\text{center} \lesssim E_\text{winding}.
\end{equation}

When we have a non-invertible symmetry, we obtain stronger bounds.
We denote the emergence scale for the non-invertible 1-form symmetry again as, by an abuse of notation, $E_\text{center}$, since it acts on the Wilson lines in the same way as the ordinary center 1-form symmetry.
In addition, we also associate to the $U(1)^{(1)}$ magnetic 1-form symmetry an emergence energy scale $E_\text{magnetic}$.
The magnetic 1-form symmetry is generated by the 2-form current $J_{\text{magnetic}}^{(2)} = \frac{1}{2} \star (qF_1)$, where $F_1$ is the field strength for $U(1)_Y$.
The fusion algebra of the non-invertible 1-form symmetry operator explained in Appendix \ref{app:symm} shows that it cannot exist without the winding 2-form symmetry and the magnetic 1-form symmetry. 
In other words, the non-invertible center 1-form symmetry is the ``child'' symmetry $G_C$, and both the magnetic 1-form symmetry and winding 2-form symmetries are ``parent'' symmetries $G_P$.
We therefore find the constraint
\begin{equation}
    E_\text{center} \lesssim \min\left\{E_\text{magnetic},E_\text{winding}\right\}.
\end{equation}

So far, we have been agnostic to the physical interpretation of the emergence scales. Let us now explore the implications for physical quantities of interest.
$E_\text{winding}$ physically corresponds to the energy scale at which axion strings become dynamical and may unwind.
This is naturally associated with the string tension, $E_\text{winding} \sim \sqrt{T}$, although it may be quite a bit smaller than $\sqrt{T}$ depending on the UV theory.
For example, in the KSVZ model considered in~\cite{Brennan:2020ehu}, it was found to be associated with $E_\text{winding}\sim (m_\rho f)^{1/2}$, where $m_\rho$ is the mass of the radial mode of the scalar field responsible for PQ symmetry breaking, and $f$ its vacuum expectation value. 
At weak couplings, $m_\rho \ll f\sim \sqrt{T}$.

$E_\text{magnetic}$ is most naturally associated with the mass of the lightest hypercharge monopole, $E_\text{magnetic} \sim m_\text{monopole}$, although depending on the UV theory, it may again be significantly smaller. 
For example, if the dynamical monopole is an 't Hooft-Polyakov monopole associated with a UV non-abelian gauge group $\mathcal{G}$, then the magnetic symmetry breaking scale is instead the Higgsing scale $v$ at which $\mathcal{G}$ is broken to $U(1)_Y$. 
The mass of the 't Hooft-Polyakov monopole is $m_\text{monopole} \sim v/g$, which is much larger than $v$ for a weakly coupled theory.

Similarly, $E_\text{center}$ can be associated with the mass of the lightest particle charged under the center of the gauge group, which we denote as $E_\text{center} \sim m_{\text{center}}$. 
To be more precise, for the case of higher-group symmetry, $m_{\text{center}}$ is the mass of the lightest particle charged under the $\mathbb{Z}_{K/q}$ subgroup of the center of the gauge group.
For the case of non-invertible symmetry, $m_{\text{center}}$ is the mass of the lightest particle charged under the $\mathbb{Z}_{6/q}$ center of the gauge group.
In both cases, we refer to $m_{\text{center}}$ loosely as the mass of the lightest ``$\mathbb{Z}_6$-charged particle.''
We emphasize that such a particle can either be a new fundamental   field  with $\mathbb{Z}_6$ gauge charge, or a  solitonic excitation of the monopole and axion string dictated by the anomaly inflow.

Combining these interpretations, we conclude that our inequalities imply
\begin{equation}
    m_\text{center} \lesssim \sqrt{T}
\end{equation}
whenever the higher-group symmetry exists, and 
\begin{equation}
    m_\text{center} \lesssim \min\left\{m_\text{monopole},\sqrt{T}\right\}
\end{equation}
whenever the non-invertible 1-form symmetry exists. 
In short, the tension of axion strings and the mass of hypercharge monopoles are generically bounded from below by the mass of the lightest particle charged under the center of the SM gauge group. 

The higher-group symmetry inequality is easy to understand for  UV completions based on the PQ mechanism. For example, consider an axion with couplings $K_i$ obeying the quantization conditions of $q=1$ in Table~\ref{table:quantization} but not $q=6$. 
Any PQ UV completion would then require at least one heavy fermion charged under the $\mathbb{Z}_{K/q}$ subgroup of the gauge group to be integrated out near the scale $f$.\footnote{
If the global form of the SM  gauge group corresponds to $q=1$ and $K_1 \in 6\mathbb{Z}$, so that we have the invertible $\mathbb{Z}_6^{(1)} \cong \mathbb{Z}_2^{(1)} \times \mathbb{Z}_3^{(1)}$ 1-form symmetry, it is possible that only the $\mathbb{Z}_2^{(1)}$ (or $\mathbb{Z}_3^{(1)}$) subgroup participates in the higher-group symmetry, depending on the values of $K_i$'s. In this case, for a heavy fermion that is neutral under the  $\mathbb{Z}_2^{(1)}$ (or $\mathbb{Z}_3^{(1)}$) subgroup of the original $\mathbb{Z}_{6}^{(1)}$, its mass would not be subject to  our inequality.
}
Therefore, $m_\text{center} \lesssim \sqrt{T} \sim f$ is automatically satisfied.
On the other hand, if the true SM gauge group had $q=1$ but the axion couplings were to satisfy the $q=6$ quantization conditions, then the higher-group symmetry would be trivial from Table~\ref{table:symmetries}, and we would not have any inequality. 
In terms of a PQ UV completion, this means that the heavy fermion(s) integrated out near the scale $f$ may have trivial charge under the $\mathbb{Z}_{K/q}$ subgroup of the gauge group, and so there is no constraint on $m_\text{center}$.
This also provides an intuitive explanation for why the higher-group symmetry conditions in Table~\ref{table:symmetries} are so similar to the quantization conditions in Table~\ref{table:quantization}, at least in the context of PQ UV completions. 
We should emphasize, however, that our inequalities apply universally to any UV completion, not restricted to the PQ mechanism.

The non-invertible symmetry  implies two inequalities, $m_\text{center} \lesssim \sqrt{T}$ as well as $m_\text{center} \lesssim m_\text{monopole}$.
The former can be  intuitively understood in PQ UV completions, similar to the higher-group case.
On the other hand,  $m_\text{center} \lesssim m_\text{monopole}$   can be  verified in models where  $U(1)_Y$ is embedded in a UV non-abelian gauge group $\mathcal{G}$.
In this case, $m_\text{center} \sim gv$ is the mass of the additional massive gauge bosons, whereas $m_\text{monopole} \sim v/g$.
The inequality $m_\text{center} \lesssim m_\text{monopole}$ is then satisfied at weak couplings. 
More generally, these inequalities can also be satisfied by the solitonic excitations on the monopoles and the axion strings from anomaly inflow (see Section 7.1 of \cite{Choi:2022fgx}).

\section{Conclusions}

The ambiguity in the global form of the SM gauge group generally has few experimentally observable effects, and is not often discussed as a result. 
However, for an axion, we have shown that this global structure is of central importance in dictating the allowed values of its quantized couplings to the SM gauge fields, resulting in the conditions in Table~\ref{table:quantization}.
These quantization conditions have immediate phenomenological implications.
In particular, the original KSVZ axion model is incompatible with both the $q=3$ and $q=6$ quantization conditions unless the heavy fermion is given a hypercharge.
For the case of $q=6$ -- the phenomenologically preferred value from the perspective of GUTs and the non-observation of cosmologically stable exotic relics -- the quantization yields a highly non-trivial correlation between $E$ and $N$. 
We have used this correlation to show that the smallest allowed effective coupling to photons $|g_{a\gamma\gamma}|$ for a post-inflationary QCD axion with domain wall number $N_\text{DW} = 1$ is realized by the ratio $E/N=8/3$. 

Of course, if an axion were discovered, other consequences would become immediately relevant.  
Unambiguously measuring all of an axion's couplings may be extremely challenging depending on the region of parameter space it populates, and would require several different experimental observables.
Nevertheless, by measuring the couplings $E$ and $N$ separately and testing the quantization conditions, one could potentially falsify $q=6$, regardless of whether the axion is of pre-inflationary or post-inflationary type and without ever discovering a $\mathbb{Z}_6$-charged particle. 
Such a discovery would greatly restrict viable UV completions of the SM gauge groups.
Additionally, generalizing~\cite{Brennan:2020ehu,Choi:2022fgx}, we have shown that there is a higher-group symmetry structure between the electric 1-form center symmetry and the $U(1)^{(2)}$ winding 2-form symmetry of the axion whenever the couplings satisfy the conditions in Table~\ref{table:symmetries}. 
Similarly, the electric 1-form symmetry also becomes non-invertible depending on the value of $E$. 
These symmetries result in model independent constraints between the masses of $\mathbb{Z}_6$-charged particles, the masses of hypercharge magnetic monopoles, and the axion string tension.

There are several avenues for further study. 
We have considered the simplest case of a single axion coupled only to the SM gauge group. 
However, in recent years, there have been a number of axion models developed going beyond these minimal assumptions~\cite{DiLuzio:2020wdo}.
A clear next step would be to generalize these results to theories of multiple axions or extended gauge groups, which may have more complicated symmetry structures and quantization conditions. 
We have also neglected discussing the 0-form shift symmetry of the axion, for good reason, since it is generically broken by the potential generated by QCD or UV instantons. 
However, it is  known that the shift symmetry can also lead to a higher-group symmetry~\cite{Hidaka:2020iaz,Hidaka:2020izy,Brennan:2020ehu,Hidaka:2021kkf,Hidaka:2021mml} or become non-invertible~\cite{Choi:2022jqy,Cordova:2022ieu,Choi:2022fgx}, which would then lead to additional inequalities. 
These may be relevant for axions that do not couple to QCD or for theories with multiple domain walls.

\section*{Acknowledgements}
We are grateful to I. Garcia Garcia,  S.\ Hong, M.\ Reece, J.\ Stout, G.\ Villadoro, and L.-T.\ Wang   for stimulating discussions. We thank A.\ Hook for comments on a draft. 
We also thank the authors of \cite{Reece:2023iqn,Agrawal:2023sbp} for coordinating submission.
Shortly after our paper appeared on arXiv, another paper \cite{Cordova:2023her} was  also posted, which  discusses the quantization of the axion-gauge couplings too. 
The work of HTL was supported in part by a Croucher fellowship from the Croucher Foundation, the Packard Foundation and the Center for Theoretical Physics at MIT. The work of SHS was supported in part by NSF grant PHY-2210182. This work was performed in part at Aspen Center for Physics during the workshop ``Traversing the Particle Physics Peaks: Phenomenology to Formal,'' which is supported by National Science Foundation grant PHY-2210452. The authors of this paper were ordered alphabetically.

\appendix

\section{More on   the higher-group and non-invertible symmetries} \label{app:symm}

In this Appendix, we provide derivations of the generalized global symmetries of the SM coupled to a single axion through the coupling \eqref{eq:axion_coupling_comp} (or equivalently, \eqref{eq:axion_coupling_diff}), which were used in the main text to obtain lower bounds on the tension of axion strings and the mass of hypercharge monopoles.
The results are summarized in Table \ref{table:symmetries}.
We work in the Euclidean signature for convenience.

\subsection{Higher-group symmetries} \label{app:highergroup}

We first derive the higher-group symmetries of the theory, following \cite{Seiberg:2018ntt,Cordova:2019uob,Hidaka:2020iaz,Hidaka:2020izy,Brennan:2020ehu}.
We will start from the $q=1$ case. 
Let us first consider an $SU(3)\times SU(2) \times U(1)_Y$ gauge theory with an axion and no other matter fields, after which we will add the SM fermions.
There are two invertible symmetries of relevance:
\begin{align*}
    &\mathbb{Z}^{(1)}_3 \times \mathbb{Z}^{(1)}_2 \times \mathbb{Z}_{|K_1|}^{(1)} ~~~ \text{electric 1-form symmetry} \,,\\
    &U(1)^{(2)} ~~~ \text{winding 2-form symmetry} \,.
\end{align*}
The electric 1-form symmetry corresponds to the product of the center symmetries of $SU(3)$ and $SU(2)$ with the $\mathbb{Z}_{|K|}^{(1)}$ electric 1-form symmetry of axion-Maxwell theory.
The charged objects are Wilson lines whose representation transforms nontrivially under the subgroup $\mathbb{Z}_3 \times \mathbb{Z}_2 \times \mathbb{Z}_{|K_1|} \subset SU(3) \times SU(2) \times U(1)_Y$.
The $U(1)^{(2)}$ winding 2-form symmetry is generated by the topological line operators $\eta^{(\text{w})}_\alpha (\gamma) \equiv \exp (i\alpha \oint_\gamma \frac{\mathrm{d}\theta}{2\pi})$, where $\alpha \sim \alpha + 2\pi$ is a $U(1)$ group parameter.
It measures the winding number of the axion field $\theta(x)$ around a closed loop $\gamma$, and the charged objects are two-dimensional worldsheets of probe axion strings.
There are also a 0-form shift symmetry and a magnetic 1-form symmetry that we will not consider, although they may lead to a higher-group symmetry as well~\cite{Brennan:2020ehu,Hidaka:2020iaz,Hidaka:2020izy,Hidaka:2021kkf,Hidaka:2021mml}.

To find whether the electric 1-form symmetry and the winding 2-form symmetry combine into a higher-group symmetry or not, one may couple the theory to the background gauge fields for these symmetries \cite{Kapustin:2013uxa,Cordova:2018cvg,Benini:2018reh}. The background gauge fields for the electric 1-form symmetry can be represented by pairs $(B_i^{(2)},\Gamma_i^{(1)})$ of $U(1)$ 2-form gauge fields $B_i^{(2)}$ and $U(1)$ 1-form gauge fields $\Gamma_i^{(1)}$ ($i=1,2,3$), satisfying \cite{Kapustin:2014gua,Gaiotto:2017yup}
\begin{equation}
    3B_3^{(2)} = \mathrm{d}\Gamma_3^{(1)} \,, \quad 2B_2^{(2)} = \mathrm{d}\Gamma_2^{(1)} \,, \quad K_1 B_1^{(2)} = \mathrm{d}\Gamma_1^{(1)}\,.
\end{equation}
The superscripts denote the form degrees of the gauge fields.
The gauge transformations of these background gauge fields are
\begin{align} \label{eq:bkgd_transf}
\begin{split}
  &B_3^{(2)} \sim B_3^{(2)} + \mathrm{d}\lambda_3^{(1)} \,, \quad \Gamma_3^{(1)} \sim \Gamma_3^{(1)} + \mathrm{d}\lambda_3^{(0)} + 3\lambda_3^{(1)} \,, \\
  &B_2^{(2)} \sim B_2^{(2)} + \mathrm{d}\lambda_2^{(1)} \,, \quad \Gamma_2^{(1)} \sim \Gamma_2^{(1)} + \mathrm{d}\lambda_2^{(0)} + 2\lambda_2^{(1)} \,, \\
  &B_1^{(2)} \sim B_1^{(2)} + \mathrm{d}\lambda_1^{(1)} \,,  \quad \Gamma_1^{(1)} \sim \Gamma_1^{(1)} + \mathrm{d}\lambda_1^{(0)} + K_1 \lambda_1^{(1)} \,.
\end{split}
\end{align}
The gauge-invariant data of the background gauge fields $(B_i^{(2)},\Gamma_i^{(1)})$ are encoded in the discrete cohomology classes which we denote as
\begin{align}
\begin{split}
    &\left[ \frac{3}{2\pi} B^{(2)}_3 \right] \equiv \widetilde{B}^{(2)}_3 \in H^2 (X, \mathbb{Z}_3) \,, \\ 
    &\left[ \frac{2}{2\pi} B^{(2)}_2 \right] \equiv \widetilde{B}^{(2)}_2 \in H^2 (X, \mathbb{Z}_2) \,, \\ 
    &\left[ \frac{K_1}{2\pi} B^{(2)}_1 \right] \equiv \widetilde{B}^{(2)}_1 \in H^2 (X, \mathbb{Z}_{|K_1|}) \,.
\end{split}
\end{align}
Here, $X$ is the spacetime manifold.
We denote the $U(1)$ 3-form background gauge field for the winding symmetry as $C^{(3)}$.

Consider coupling the theory only to the electric 1-form symmetry background gauge fields, but not the winding symmetry background gauge field. 
The Lagrangian after naively coupling to these background gauge fields is given by
\begin{align} \label{eq:naive_q=1}
  \begin{split}
    &\frac{f^2}{2} \mathrm{d}\theta \wedge \star \mathrm{d}\theta + \frac{1}{g_3^2} \text{Tr}\, F_3 \wedge \star F_3 + \frac{1}{g_2^2} \text{Tr}\, F_2 \wedge \star F_2 + \frac{1}{2g_1^2} (F_1 - B_1^{(2)})\wedge \star (F_1 - B_1^{(2)}) \\ \quad & - 
    \frac{iK_3}{8\pi^2} \theta \, \text{Tr}\, F_3 \wedge F_3 - 
    \frac{iK_2}{8\pi^2} \theta \, \text{Tr}\, F_2 \wedge F_2 -
    \frac{iK_1}{8\pi^2} \theta \, (F_1 - B_1^{(2)}) \wedge (F_1 - B_1^{(2)}) \,,
  \end{split}
  \end{align}
where it is understood that $A_3$ now is a $PSU(3)$ gauge field with $w_2(A_3) = \widetilde{B}^{(2)}_3$, and similarly $A_2$ is a $PSU(2) = SO(3)$ gauge field with $w_2(A_2)= \widetilde{B}^{(2)}_2$ \cite{Kapustin:2014gua,Seiberg:2018ntt}.
However, with the naive coupling \eqref{eq:naive_q=1}, the exponentiated action generally does not respect the $2\pi$ periodicity of $\theta$.
Rather, under $\theta \sim \theta + 2\pi$, it picks up \cite{Aharony:2013hda,Kapustin:2014gua,Gaiotto:2014kfa,Hsin:2018vcg,Cordova:2019uob,Hsin:2020nts}
\begin{align} \label{eq:q1_phase}
\begin{split}
  &\text{exp}\left(\frac{i}{4\pi} 6 K_3 \int B^{(2)}_3 \wedge B^{(2)}_3  +
   \frac{i}{4\pi} 2 K_2 \int B^{(2)}_2 \wedge B^{(2)}_2 
   + \frac{i}{4\pi} K_1 \int B_1^{(2)} \wedge B_1^{(2)}\right) \,.
\end{split}
\end{align}
The coupling \eqref{eq:naive_q=1} is inconsistent unless the phase \eqref{eq:q1_phase} is trivial, which is the case if and only if $K_3 \in 3\mathbb{Z}$, $K_2 \in 2\mathbb{Z}$, and $|K_1| = 0,1$.
If these conditions are satisfied, we do not find any nontrivial higher-group symmetry (see below).

When \eqref{eq:q1_phase} is not trivial, to consistently couple to the background gauge fields for the electric 1-form symmetry, we are forced to also couple to the background gauge field for the winding 2-form symmetry at the same time, such that the Lagrangian contains the term $\frac{i}{2\pi}\theta G^{(4)}$, where $G^{(4)}$ is the invariant field strength given by
\begin{equation} \label{eq:q=1_inv_field_strength}
  G^{(4)} = \mathrm{d}C^{(3)} + \frac{1}{4\pi} \left(
    6K_3 B_3^{(2)} \wedge B_3^{(2)} + 2K_2 B_2^{(2)} \wedge B_2^{(2)} + K_1 B_1^{(2)} \wedge B_1^{(2)}
  \right) \,.
\end{equation}
For such a coupling to be gauge-invariant under \eqref{eq:bkgd_transf}, $C^{(3)}$ must transform as
\begin{align} \label{eq:q1_C_transform}
\begin{split}
  C^{(3)} \sim C^{(3)} &- \frac{6K_3}{2\pi} \lambda_3^{(1)} B_3^{(2)} - \frac{6K_3}{4\pi} \lambda_3^{(1)} \mathrm{d}\lambda_3^{(1)} - \frac{2K_2}{2\pi} \lambda_2^{(1)} B_2^{(2)} - \frac{2K_2}{4\pi} \lambda_2^{(1)} \mathrm{d}\lambda_2^{(1)} \\
  & - \frac{K_1}{2\pi} \lambda_1^{(1)} B_1^{(2)} - \frac{K_1}{4\pi} \lambda_1^{(1)} \mathrm{d}\lambda_1^{(1)} \,.
\end{split}
\end{align} 
This shows that there is a higher group symmetry formed by the electric 1-form symmetry and the winding 2-form symmetry, where the former is the ``child'' symmetry that cannot exist without the latter ``parent'' symmetry.
To be more precise, whether the higher-group characterized by the invariant field strength \eqref{eq:q=1_inv_field_strength} is trivial or not (that is, whether the higher-group is split or not), depends on the values of $K_i$'s as mentioned above.
For instance, suppose $K_3 = 3m$, where $m$ is an integer.
We may rewrite the invariant field strength \eqref{eq:q=1_inv_field_strength} as
\begin{equation}
  G^{(4)} = \mathrm{d}\left(C^{(3)} + \frac{2m}{4\pi} \Gamma_3^{(1)} d \Gamma_3^{(1)}\right) + \frac{1}{4\pi} \left(
   2K_2 B_2^{(2)} \wedge B_2^{(2)} + K_1 B_1^{(2)} \wedge B_1^{(2)}\right) \,.
\end{equation}
We can then redefine the $U(1)^{(2)}$ winding 2-form symmetry background gauge field by\footnote{Mathematically, both the $U(1)$ 2-form gauge field $C^{(3)}$ and the Chern-Simons term $\frac{2m}{4\pi} \Gamma_3^{(1)}\mathrm{d}\Gamma_3^{(1)}$ ought to be thought of as elements in the degree-4 differential cohomology group $\check{H}^4(X)$, and it makes sense to shift $C^{(3)}$ by $\frac{2m}{4\pi} \Gamma_3^{(1)}\mathrm{d}\Gamma_3^{(1)}$ since $\check{H}^4(X)$ is an abelian group. See, for instance, \cite{Hsieh:2020jpj}.} $C^{(3)} \rightarrow C'^{(3)} \equiv C^{(3)} + \frac{2m}{4\pi} \Gamma_3^{(1)} \mathrm{d} \Gamma_3^{(1)}$ and see that the $\mathbb{Z}_3^{(1)}$ subgroup of the electric 1-form symmetry does not participate in the higher group.
By similar arguments, one can show that the $\mathbb{Z}_2^{(1)}$ subgroup of the electric 1-form symmetry participates nontrivially in the higher-group if and only if $K_2 \notin 2\mathbb{Z}$, whereas the $\mathbb{Z}_{|K_1|}^{(1)}$ subgroup always participates nontrivially (unless $|K_1| = 0,1$ and the group $\mathbb{Z}_{|K_1|}^{(1)}$ itself is trivial).
Note that $K_3 \in 3\mathbb{Z}$, $K_2 \in 2\mathbb{Z}$, and $|K_1| =0,1$ are precisely the conditions that the phase \eqref{eq:q1_phase} is trivial, in which case there is no nontrivial higher-group symmetry at all.

Anticipating adding fermions, let us combine the background gauge fields $\widetilde{B}_3^{(2)}$ and $\widetilde{B}_2^{(2)}$ into a single $\mathbb{Z}_6^{(1)}$ background gauge field (recall $\mathbb{Z}_6 \cong \mathbb{Z}_2 \times \mathbb{Z}_3$), described by a pair $(B_6^{(2)},\Gamma_6^{(1)})$ of $U(1)$ 2-form and 1-form gauge fields satisfying
\begin{equation}
      6B_6^{(2)} = \mathrm{d}\Gamma_6^{(1)} \,.
\end{equation}
The gauge-invariant data are encoded into the $\mathbb{Z}_6$ cohomology class $ \left[ \frac{6}{2\pi} B^{(2)}_6 \right] \equiv \widetilde{B}^{(2)}_6 \in H^2 (X, \mathbb{Z}_6)$,
which decomposes into the original background gauge fields as
\begin{equation} \label{eq:combine_Z6}
  \widetilde{B}_6^{(2)} = 2\widetilde{B}_3^{(2)} + 3\widetilde{B}_2^{(2)} \,.
\end{equation}
Mathematically, the 2 in front of $\widetilde{B}_3^{(2)}$ stands for the map $2: H^2(X,\mathbb{Z}_3) \rightarrow H^2(X,\mathbb{Z}_6)$ coming from the homomorphism $\mathbb{Z}_3 \xrightarrow{\times 2} \mathbb{Z}_6$, and similarly the 3 in front of $\widetilde{B}_2^{(2)}$ stands for the map $3: H^2(X,\mathbb{Z}_2) \rightarrow H^2(X,\mathbb{Z}_6)$ coming from $\mathbb{Z}_2 \xrightarrow{\times 3} \mathbb{Z}_6$.
Physically, \eqref{eq:combine_Z6} can be understood by checking that the Wilson surface for $\widetilde{B}_6^{(2)}$ correctly factorizes into the Wilson surfaces of $\widetilde{B}_3^{(2)}$ and $\widetilde{B}_2^{(2)}$, $\exp(\frac{2\pi i}{6} \oint \widetilde{B}_6^{(2)}) = \exp(\frac{2\pi i}{3} \oint \widetilde{B}_3^{(2)})\exp(\frac{2\pi i}{2} \oint \widetilde{B}_2^{(2)})$. 
In terms of the continuous fields $(B_i^{(2)},\Gamma_i^{(2)})$, the relation \eqref{eq:combine_Z6} translates into
\begin{equation}
  B_6^{(2)} = B_3^{(2)} + B_2^{(2)} \,, \quad
  \Gamma_6^{(1)} = 2\Gamma_3^{(1)} + 3\Gamma_2^{(1)} \,.
\end{equation}
The invariant field strength $G^{(4)}$ in terms of the $\mathbb{Z}_6^{(1)}$ gauge field is given by
\begin{equation}
  G^{(4)} = \mathrm{d}C^{(3)} + \frac{1}{4\pi} \left(
    6k_6 B_6^{(2)} \wedge B_6^{(2)} + K_1 B_1^{(2)} \wedge B_1^{(2)}
  \right) \,,
\end{equation}
where $k_6$ is a combination of $K_2$ and $K_3$ defined modulo 6 that we must now determine. 
In order to find the dependence of $k_6$ on $K_2$ and $K_3$, we insert $B_6^{(2)} = B_3^{(2)} + B_2^{(2)}$ to the invariant field strength. 
This gives, modulo shifts in $C^{(3)}$, 
\begin{equation}
  G^{(4)} = \mathrm{d}C^{(3)}+ \frac{1}{4\pi} \left(
      3(2k_6) B_3^{(2)} \wedge B_3^{(2)} + 
      2(3k_6) B_2^{(2)} \wedge B_2^{(2)} + 
      K_1  B_1^{(2)} \wedge B_1^{(2)}
  \right) \,.
\end{equation}
By comparing this expression with \eqref{eq:q=1_inv_field_strength} we see that
\begin{equation}
  2K_3 = 2k_6 \quad \text{mod 3} \quad \text{and} \quad
  K_2 = 3k_6 \quad \text{mod 2} \,,
\end{equation}
which is equivalent to\footnote{We first invert the equations as $k_6 = 2^{-1}(2K_3) = 2(2K_3)$ mod 3 and $k_6 = 3^{-1}K_2 = K_2$ mod 2. The value of $k_6$ mod 6 is uniquely determined due to the Chinese remainder theorem, and we can explicitly check that $k_6 = 4K_3 + 3K_2$ mod 6 is a solution.}
\begin{equation}
  k_6 =  4K_3 + 3K_2 \quad \text{mod 6} \,.
\end{equation}
Therefore, the $\mathbb{Z}_6^{(1)}$ 1-form symmetry participates nontrivially in the higher group if and only if $4K_3 + 3K_2 \neq 0$ mod 6, and the final invariant field strength may be written as
\begin{equation} \label{eq:q=1_inv_field_strength_Z6}
  G^{(4)} = \mathrm{d}C^{(3)} + \frac{1}{4\pi} \left(
    6(4K_3 + 3K_2) B_6^{(2)} \wedge B_6^{(2)} + K_1 B_1^{(2)} \wedge B_1^{(2)}
  \right) \,.
\end{equation}

We are finally in a position to consider the addition of the SM matter fields. 
The preserved (invertible) electric 1-form symmetry is given by a subgroup $\mathbb{Z}^{(1)}_{K} \subset \mathbb{Z}_6^{(1)} \times \mathbb{Z}^{(1)}_{|K_1|}$, where we have defined $K \equiv \text{gcd}(6,K_1)$. 
On the other hand, the $U(1)^{(2)}$ winding 2-form symmetry is unaffected by the SM matter fields.
We denote the background gauge field for the $\mathbb{Z}_K^{(1)}$ symmetry by the pair $(B^{(2)},\Gamma^{(1)})$ of $U(1)$ 2-form and 1-form gauge fields satisfying
\begin{equation}
  K B^{(2)} = \mathrm{d}\Gamma^{(1)} \,.
\end{equation}
Turning on the background gauge field $B^{(2)}$ for the $\mathbb{Z}_K^{(1)}$ subgroup which is preserved after the coupling to the SM matter fields amounts to setting
\begin{equation}
  B_6^{(2)} = B_1^{(2)} = B^{(2)} \,.
\end{equation}
The invariant field strength for the higher-group symmetry which mixes the $\mathbb{Z}_K^{(1)}$ electric 1-form symmetry and the $U(1)^{(2)}$ winding 2-form symmetry becomes
\begin{equation} \label{eq:q=1_inv_field_strength_SM}
  G^{(4)} = \mathrm{d}C^{(3)} + \frac{k}{4\pi} K B^{(2)} \wedge B^{(2)}  \,,
\end{equation}
where $k$ is a combination of $(K_3,K_2,K_1)$ to be determined shortly.
The value of $k$ is meaningful only modulo $K$, since shifting $C^{(3)}$ by the Chern-Simons term $\frac{1}{4\pi} \Gamma^{(1)} \mathrm{d} \Gamma^{(1)}$ can change the value of $k$ by any integer multiple of $K$.
By comparison with \eqref{eq:q=1_inv_field_strength_Z6} we immediately find
\begin{equation}
  q=1:~~~ k =\dfrac{24K_3 + 18K_2 + K_1}{K} \quad \text{mod $K$} \,.
\end{equation}
Therefore, for $q=1$, a nontrivial higher-group symmetry survives the coupling to the SM matter fields if and only if $k \neq 0$ mod $K$.
Note that the higher group can still be nontrivial even if $K_1 = 0$ mod 6, that is, when there is no non-invertible 1-form symmetry (see Appendix~\ref{sec:non-invertible}).
Whenever the higher-group symmetry is nontrivial (i.e., is not a split higher-group), the ``child'' $\mathbb{Z}^{(1)}_{K}$ electric 1-form symmetry cannot exist without the ``parent'' $U(1)^{(2)}$ winding 2-form symmetry.

The higher-group symmetry for the other values of $q$ can also be obtained analogously.
Consider the $q=2$ case.
Similar to before, we begin with the $[SU(3)\times SU(2) \times U(1)_Y]/\mathbb{Z}_2 = SU(3) \times U(2)$ gauge theory coupled to a single axion without the SM matter fields. 
The global symmetries of interest are the (invertible) $\mathbb{Z}_3^{(1)} \times \mathbb{Z}^{(1)}_{|K_1|/2}$ electric 1-form symmetry, 
as well as the $U(1)^{(2)}$ winding 2-form symmetry as before.
Note that the quantization condition of $K_i$'s for $q=2$ requires $K_1 \in 2\mathbb{Z}$, and hence the $|K_1|/2$ in $\mathbb{Z}^{(1)}_{|K_1|/2}$ is a nonnegative integer.
The $\mathbb{Z}_3^{(1)}$ part of the electric 1-form symmetry is the usual center 1-form symmetry for the $SU(3)$ gauge field.
On the other hand, the $\mathbb{Z}^{(1)}_{|K_1|/2}$ 1-form symmetry is generated by the topological surface operator (see, for instance, \cite[Appendix A]{Choi:2022fgx} and references therein)
\begin{equation} \label{eq:q2_ZK2}
  \exp \left[\frac{2\pi i}{(K_1/2)} \oint_\Sigma \left(
    -\frac{i}{e^2} \star F_1 
    -\frac{(K_1/2)}{4\pi^2} \theta (2F_1)
  \right)\right] \,.
\end{equation}
The operator \eqref{eq:q2_ZK2} is topological due to the equation of motion, and moreover is gauge-invariant under the identification $\theta \sim \theta + 2\pi$ since $\oint F_1 \in \pi \mathbb{Z}$ for $q=2$ as was discussed in Section \ref{sec:fractional}.

We may now couple the theory to the background gauge fields for the electric 1-form symmetry as well as the winding 2-form symmetry.
The background gauge fields for the $\mathbb{Z}_3^{(1)} \times \mathbb{Z}_{|K_1|/2}^{(1)}$ electric 1-form symmetry are pairs $(B_3^{(2)},\Gamma_3^{(1)})$, $(B_1^{(2)},\Gamma_1^{(1)})$ of $U(1)$ 2-form and 1-form gauge fields satisfying $3B_3^{(2)} = \mathrm{d}\Gamma_3^{(1)}$ and $\frac{K_1}{2} B_1^{(2)} = \mathrm{d}\Gamma_1^{(1)}$.
Consider coupling the theory only to these electric 1-form symmetry background fields.
In the presence of them, the exponentiated action is not gauge-invariant under $\theta \sim \theta + 2\pi$, but picks up \cite{Aharony:2013hda,Kapustin:2014gua,Gaiotto:2014kfa,Hsin:2018vcg,Cordova:2019uob,Hsin:2020nts}
\begin{align} \label{eq:q2_phase}
\begin{split}
  &  \exp\left(\frac{i}{4\pi} 6K_3 \int B_3^{(2)} \wedge B_3^{(2)}  + \frac{i}{4\pi} K_1\int B_1^{(2)} \wedge B_1^{(2)}
   + \frac{i}{4\pi} (2K_2 + K_1) \int F_1 \wedge F_1 \right) \\
  &= \exp\left(\frac{i}{4\pi} 6K_3 \int B_3^{(2)} \wedge B_3^{(2)}  + \frac{i}{4\pi} K_1 \int B_1^{(2)} \wedge B_1^{(2)}
  \right) \,,
\end{split}
\end{align}
where we have used the quantization condition $2K_2 + K_1 \in 4\mathbb{Z}$ and $\frac{1}{2}\int c_1(F_{U(2)}) \wedge c_1(F_{U(2)}) = \frac{1}{2\pi^2} \int F_1 \wedge F_1 \in \mathbb{Z}$ on spin manifolds, both of which hold for $q=2$.
Unless $K_3 \in 3\mathbb{Z}$ and $|K_1|/2 = 0,1$, the phase \eqref{eq:q2_phase} is nontrivial.
Similar to before, this implies that to consistently couple the theory to the electric 1-form symmetry background gauge fields, we are forced to introduce the background gauge field $C^{(3)}$ for the $U(1)^{(2)}$ winding 2-form symmetry by the term $\frac{i}{2\pi}\theta G^{(4)}$ in the Lagrangian, where the invariant field strength $G^{(4)}$ is given by
\begin{equation}
  G^{(4)} = \mathrm{d}C^{(3)} + \frac{1}{4\pi} \left(
    6K_3 B_3^{(2)} \wedge B_3^{(2)} + 2\left(\frac{K_1}{2}\right) B_1^{(2)} \wedge B_1^{(2)}
  \right) \,.
\end{equation}
This characterizes a nontrivial higher-group symmetry (for values of $K_3$ and $K_1$ for which \eqref{eq:q2_phase} is nontrivial, namely $K_3 \notin 3\mathbb{Z}$ or $|K_1/2| \neq 0,1$), since now $C^{(3)}$ must transform under gauge transformations of the electric 1-form symmetry background gauge fields, analogous to \eqref{eq:q1_C_transform}.

Introducing the SM fermions breaks the $\mathbb{Z}_3^{(1)} \times \mathbb{Z}_{|K_1|/2}^{(1)}$ 1-form symmetry down to its $\mathbb{Z}_{K/2}^{(1)} = \mathbb{Z}_{\text{gcd}(3,K_1/2)}^{(1)}$ subgroup, where $K \equiv \text{gcd}(6,K_1) \in 2\mathbb{Z}$.
The invariant field strength for the higher-group symmetry is obtained by setting $B_3^{(2)} = B_1^{(2)} = B^{(2)}$ where the pair $(B^{(2)},\Gamma^{(1)})$, satisfying $\frac{K}{2}B^{(2)} = d\Gamma^{(1)}$, is the $\mathbb{Z}_{K/2}^{(1)}$ background gauge field.
This gives
\begin{equation}
  G^{(4)} = \mathrm{d}C^{(3)} + \frac{k}{4\pi} \left(\frac{K}{2}\right) B^{(2)} \wedge B^{(2)} \,,
\end{equation}
where
\begin{equation}
  q=2:~~~ k = \frac{6K_3 + K_1}{K/2}  \quad \text{mod $\frac{K}{2}$}
\end{equation}
determines the existence of the nontrivial higher-group symmetry for the $q=2$ case.
When $k$ is nonzero, the ``child'' $\mathbb{Z}_{K/2}^{(1)}$ electric 1-form symmetry cannot exist without the ``parent'' $U(1)^{(2)}$ winding 2-form symmetry. 

The $q=3$ case follows the same steps, and we will be brief. 
The gauge group is $[SU(3)\times SU(2) \times U(1)_Y]/\mathbb{Z}_3 = SU(2)\times U(3)$.
If we consider an $SU(2)\times U(3)$ gauge theory coupled to an axion, there is an invertible $\mathbb{Z}_2^{(1)} \times \mathbb{Z}^{(1)}_{|K_1|/3}$ electric 1-form symmetry as well as the $U(1)^{(2)}$ winding 2-form symmetry. 
The addition of the SM matter fields breaks the former down to its $\mathbb{Z}_{K/3}^{(1)} = \mathbb{Z}_{\text{gcd}(2,K_1/3)}^{(1)}$ subgroup (recall that $K/3 \in \mathbb{Z}$ if $q=3$ due to the quantization condition on $K_i$'s). 
One finds a nontrivial higher-group symmetry mixing the $\mathbb{Z}_{K/3}^{(1)}$ electric 1-form symmetry with the $U(1)^{(2)}$ winding 2-form symmetry if and only if
\begin{equation}
  q=3:~~~ k = \frac{2K_2 + K_1}{K/3}  \quad \text{mod $\frac{K}{3}$}
\end{equation}
is nonzero.
When $k$ is nonzero, the ``child'' $\mathbb{Z}_{K/3}^{(1)}$ electric 1-form symmetry cannot exist without the ``parent'' $U(1)^{(2)}$ winding 2-form symmetry. 

For $q=6$, there is no electric 1-form symmetry in the SM, and there is no interesting higher-group symmetry to be discussed.

Finally, we briefly mention that the quantization conditions on $K_i$'s discussed in Section \ref{sec:quantization} can also be derived from the higher-group symmetry structure for the $q=1$ case.
The idea is that the gauge groups with $q>1$ can be realized by first starting from the SM coupled to an axion, based on the $q=1$ gauge group $SU(3)\times SU(2) \times U(1)_Y$, and then gauging a $\mathbb{Z}_q^{(1)} \subset \mathbb{Z}_{|K_1|}^{(1)}$ subgroup of the center 1-form symmetry.
For this to make sense, first of all, we need $K_1 \in q\mathbb{Z}$, which is indeed a part of the quantization condition.
Second, to be able to actually gauge the $\mathbb{Z}_q^{(1)} \subset \mathbb{Z}_{|K_1|}^{(1)}$ subgroup of the center 1-form symmetry, the subgroup must participate trivially in the higher-group.
After straightforward manipulations on the invariant field strength \eqref{eq:q=1_inv_field_strength_SM}, one can show that this recovers the quantization conditions given in Table \ref{table:quantization}.

\subsection{Non-invertible symmetries}\label{sec:non-invertible}

In the presence of an axion, there are two classes of non-invertible global symmetries that have been considered in the literature so far.
One is the non-invertible 0-form symmetry which shifts the axion field \cite{Cordova:2022ieu,Choi:2022fgx}, and the other is the non-invertible 1-form symmetry which originates from the center symmetry in the absence of the axion \cite{Choi:2022fgx,Yokokura:2022alv}.
Here, we demonstrate the existence of the non-invertible 1-form global symmetry in the SM coupled to an axion.
The analysis is a generalization of the one provided in \cite{Choi:2022fgx}.

We begin from the $q=1$ case.
Consider first a pure $SU(3) \times SU(2) \times U(1)_Y$ gauge theory which is not coupled to any matter fields nor the axion field.
The theory has the center 1-form symmetry $\mathbb{Z}_3^{(1)} \times \mathbb{Z}_2^{(1)} \times U(1)^{(1)}$.
They are generated by topological surface operators \cite{Gaiotto:2014kfa}, and we denote them respectively as $\eta_3(\Sigma)$, $\eta_2(\Sigma)$, and $U_\alpha (\Sigma) = \exp\left[i\alpha \oint_\Sigma \left(-\frac{i}{e^2} \star F_1 \right)  \right]$.
The operator $U_\alpha (\Sigma)$ is topological due to the equation of motion $\mathrm{d}\star F_1 = 0$, and this fact plays an important role below.
The $U(1)$ group parameter $\alpha$ is normalized to be $2\pi$-periodic.

In the SM (without an axion), the center 1-form symmetry is broken down to a $\mathbb{Z}_6^{(1)}$ subgroup.
The topological surface operator which generates the $\mathbb{Z}_6^{(1)}$ symmetry is\footnote{To be more precise, the presence of the SM matter fields further modifies the expression for the symmetry operator $U(\Sigma)$, but it will not be important for our discussions.}
\begin{equation} \label{eq:q1_U}
    U(\Sigma) \equiv \eta_3(\Sigma) \eta_2(\Sigma) U_{2\pi/6} (\Sigma) \,.
\end{equation}
Once we further couple the theory to an axion as in \eqref{eq:axion_coupling_diff}, $U(\Sigma)$ ceases to be topological if $K_1 \neq 0$ due to the modification of the equation of motion, $\mathrm{d} \left(-\frac{i}{e^2} \star F\right) = \frac{K_1}{4\pi^2} \mathrm{d}\theta F_1 + \cdots$.\footnote{On the other hand, the couplings $K_2$ and $K_3$ between the axion and the nonabelian gauge fields do not affect the topological nature of the center symmetry operator, although they can lead to a nontrivial higher-group structure discussed in the previous subsection.}
Instead, one considers a new operator
\begin{equation} \label{eq:hatU_q1}
    `` \,\hat{U}(\Sigma) \equiv U(\Sigma)\exp \left[ \frac{2\pi i}{6} \oint_\Sigma \left(- \frac{K_1}{4\pi^2} \theta F_1 \right)\right] " 
\end{equation}
where the exponential term is there to compensate the effect of the axion coupling to the gauge field $A_1$.
The new operator $\hat{U}(\Sigma)$ is topological, although it is generally not a gauge-invariant operator under the identification $\theta \sim \theta + 2\pi$.
We put the quotation marks in \eqref{eq:hatU_q1} to emphasize this fact, and also similarly below for the expressions which are potentially ill-defined and/or not gauge-invariant.
To be more precise, $\hat{U}(\Sigma)$ is gauge-invariant if and only if $K_1 = 0$ mod 6, in which case the $\mathbb{Z}_6^{(1)}$ symmetry (now generated by $\hat{U}(\Sigma)$) is unbroken and there is no non-invertible 1-form symmetry.\footnote{To be more precise, there is no non-invertible 1-form symmetry operator which is not factorized into an invertible 1-form symmetry operator and a condensation operator.}

Now suppose $K_1 \neq 0$ mod 6.
As before, we write $K \equiv \text{gcd}(6,K_1)$.
Let $n \equiv \frac{6}{K} \in \mathbb{Z}$ and $m \equiv \frac{K_1}{K} \in \mathbb{Z}$, which are coprime to each other, $\text{gcd}(n,m) =1$.
Instead of the ill-defined, non-gauge-invariant operator $\hat{U}(\Sigma)$, we define the following surface operator which is gauge-invariant:\footnote{More generally, one may consider non-invertible symmetry operators which replace the non-gauge-invariant operator $\hat{U}(\Sigma)^\ell$ with $\ell$ not an integer multiple of $6/K$. We will focus on $\ell=1$ for simplicity.}
\begin{equation} \label{eq:noninv_q1}
    q=1:~~~ \mathcal{D}(\Sigma) \equiv U(\Sigma) \int [D\phi Dc]_\Sigma   
    \exp \left[ i
        \oint_\Sigma \left( 
            \frac{n}{2\pi} \phi \mathrm{d}c + \frac{m}{2\pi} \theta \mathrm{d}c + \frac{1}{2\pi} \phi F_1 
        \right)
    \right] \,.
\end{equation}
Here, $\phi \sim \phi + 2\pi$ is a compact scalar field and $c$ is a $U(1)$ gauge field, and they are auxiliary fields living on the support $\Sigma$ of the operator $\mathcal{D}(\Sigma)$.
Intuitively, \eqref{eq:noninv_q1} can be viewed as the naive symmetry operator $U(\Sigma)$ dressed with a 2d $\mathbb{Z}_n$ gauge theory (described by the $\phi$ and $c$ fields \cite{Banks:2010zn,Kapustin:2014gua}) coupled to the bulk $\theta$ and $A_1$ fields, whose effective response action is given by ``$\exp\left[ \frac{2\pi i}{6} \oint_\Sigma \left( - \frac{K_1}{4\pi^2} \theta F_1 \right) \right]$.''
The equation of motion for the $c$ gauge field gives ``$\phi = -\frac{m}{n} \theta = -\frac{K_1}{6} \theta$,'' which is not well-defined unless $K_1 = 0$ mod 6 since both $\phi$ and $\theta$ are $2\pi$-periodic scalar fields.
Naively plugging the equation of motion back to \eqref{eq:noninv_q1} recovers \eqref{eq:hatU_q1}.

The surface operator $\mathcal{D}(\Sigma)$ is topological, which can be heuristically understood from its relation to \eqref{eq:hatU_q1}.
For a more rigorous proof, see \cite{Choi:2022fgx}.
Moreover, it obeys a non-invertible fusion algebra,
\begin{equation} \label{eq:q1_fusion}
    \mathcal{D}(\Sigma) \times \overline{\mathcal{D}}(\Sigma) = \frac{1}{|H^0(\Sigma,\mathbb{Z}_n)|} \left[ 
        \sum_{i=1}^n  \left( \eta^{(\text{m})}_{2\pi/n} (\Sigma) \right)^i
    \right] \times \left[ 
        \sum_{\gamma \in H_1 (\Sigma,\mathbb{Z}_n)} \eta^{(\text{w})}_{2\pi m/n}(\gamma)
    \right] \,.
\end{equation}
Here, $\overline{\mathcal{D}}(\Sigma)$ stands for the orientation-reversal of $\mathcal{D}(\Sigma)$, $\eta^{(\text{m})}_{\alpha}(\Sigma) = \exp (i\alpha \oint_\Sigma \frac{F_1}{2\pi})$ is the $U(1)^{(1)}$ magnetic 1-form global symmetry operator, and $\eta^{(\text{w})}_{\alpha}(\gamma) = \exp (i\alpha \oint_\gamma \frac{\mathrm{d}\theta}{2\pi})$ is the $U(1)^{(2)}$ winding 2-form global symmetry operator.
The $U(1)$ group parameter $\alpha$ is always normalized such that it is $2\pi$-periodic.
The right-hand side of \eqref{eq:q1_fusion} is an example of a condensation operator \cite{Roumpedakis:2022aik,Choi:2022zal}.
The fusion algebra \eqref{eq:q1_fusion} demonstrates that the non-invertible 1-form symmetry generated by $\mathcal{D}$ is subordinate to the magnetic as well as winding symmetries.
Namely, the non-invertible 1-form symmetry (``child'') for $K_1 \neq 0$ mod 6 cannot exist without the other two invertible higher-form symmetries (``parents'').

We now move on to the $q=2$ case.
It is again convenient to start from a pure $[SU(3)\times SU(2)\times U(1)_Y]/\mathbb{Z}_2 = SU(3) \times U(2)$ gauge theory without matter fields.
The center 1-form symmetry is $\mathbb{Z}_3^{(1)} \times U(1)^{(1)}$, generated by $\eta_3(\Sigma)$ and $U_\alpha (\Sigma) = \exp\left[i\alpha \oint_\Sigma \left(-\frac{i}{e^2} \star F_1 \right)  \right]$.
The latter symmetry measures the hypercharge of the $U(2)$ Wilson lines, and we have $\alpha \sim \alpha + 2\pi$.

When we couple to the SM matter fields, the center symmetry is broken down to a $\mathbb{Z}_3^{(1)}$ subgroup which is generated by
\begin{equation}
    U(\Sigma) \equiv \eta_3 (\Sigma) U_{2\pi/3}(\Sigma) \,,
\end{equation}
where we are using the same symbol $U(\Sigma)$ as in \eqref{eq:q1_U} by an abuse of notation.
When we further couples the theory to an axion with $K_1 \neq 0$, $U(\Sigma)$ becomes non-topological similar to the $q=1$ case.
One may consider
\begin{equation} \label{eq:q2_hatU}
    `` \, \hat{U}(\Sigma) \equiv U(\Sigma) \exp \left[\frac{2\pi i}{3} \oint \left( 
        -\frac{(K_1/2)}{4\pi^2} \theta (2F_1)
    \right) \right] . \,"
\end{equation}
We have written \eqref{eq:q2_hatU} in a way that emphasizes the fact that $\frac{K_1}{2} \in \mathbb{Z}$ and $\oint \frac{2F_1}{2\pi} \in \mathbb{Z}$ for $q=2$.
Unless $\frac{K_1}{2} = 0$ mod 3, the operator $\hat{U}(\Sigma)$ is not gauge-invariant under the identification $\theta \sim \theta +2\pi$.
On the other hand, if $\frac{K_1}{2} = 0$ mod 3, $\hat{U}(\Sigma)$ is well-defined and generates an invertible $\mathbb{Z}_3^{(1)}$ 1-form symmetry of the theory.

When $\frac{K_1}{2} \neq 0$ mod 3, we instead have a non-invertible 1-form symmetry, generated by the topological surface operator
\begin{equation} \label{eq:noninv_q2}
    q=2:~~~ \mathcal{D}(\Sigma) \equiv U(\Sigma) \int [D\phi Dc]_\Sigma   
    \exp \left[ i
        \oint_\Sigma \left( 
            \frac{n}{2\pi} \phi \mathrm{d}c + \frac{m}{2\pi} \theta \mathrm{d}c + \frac{1}{2\pi} \phi (2F_1) 
        \right)
    \right] \,.
\end{equation}
It satisfies a non-invertible fusion algebra similar to \eqref{eq:q1_fusion}, which shows that the ``child'' non-invertible 1-form symmetry cannot exist without the ``parents'' magnetic 1-form symmetry and the winding 2-form symmetry.

The analysis is the same for the $q=3$ case and we will only mention the result.
For $q=3$, we have $K_1 \in 3\mathbb{Z}$, and the non-invertible 1-form symmetry exists when $\frac{K_1}{3} \neq 0$ mod 2.
It is generated by
\begin{equation} \label{eq:noninv_q3}
    q=3:~~~ \mathcal{D}(\Sigma) \equiv U(\Sigma) \int [D\phi Dc]_\Sigma   
    \exp \left[ i
        \oint_\Sigma \left( 
            \frac{n}{2\pi} \phi \mathrm{d}c + \frac{m}{2\pi} \theta \mathrm{d}c + \frac{1}{2\pi} \phi (3F_1) 
        \right)
    \right] \,,
\end{equation}
where now $U(\Sigma) \equiv \eta_2(\Sigma) U_{\pi}(\Sigma)$.
The ``child'' non-invertible 1-form symmetry again cannot exist without the two ``parent'' invertible higher-form symmetries.

For $q=6$, there is no interesting non-invertible 1-form symmetry operator, since we do not have a center 1-form symmetry in the SM to begin with even in the absence of an axion.

\section{Fractional axion-gauge couplings}\label{app:fractionalK}

Throughout the paper we assume that there are no other degrees of freedom in addition to the SM fields and a single axion field. 
However, certain topological degrees of freedom can result in a fractional axion-gauge coupling \cite{Seiberg:2010qd}, which we review below. 
It would be interesting to further explore phenomenological consequences of this mechanism.

We demonstrate this phenomenon for a dynamical axion field $\theta(x)$ coupled to  an $SU(N)$ gauge field $A$ with coupling ${K\over 8\pi^2} \theta\, \text{Tr}F\wedge F$ with $K\in \mathbb{Z}$. 
We introduce two new dynamical fields, a  periodic scalar field $\hat \theta(x)\sim \hat \theta(x)+2\pi$, and a  3-form gauge field $\mathsf{c}$. We consider the following couplings:
\begin{equation}\label{fractionalK}
{K\over 8\pi^2} \theta\, \text{Tr}F\wedge F
+{1\over 2\pi}(\hat\theta - N\theta) \mathrm{d}\mathsf{c} \,.
\end{equation}
Locally, we can integrate out $\mathsf{c}$, and obtain 
\begin{equation}\label{hattheta}
    ``\theta=\hat\theta/N ."
\end{equation}
Substituting this back into \eqref{fractionalK}, we find an  fractional axion-gauge coupling $K/N$ for $\hat\theta$. 
Globally, however, we cannot integrate out $\mathsf{c}$ and relate two compact scalar fields of the same period $2\pi$ as in \eqref{hattheta} (and hence the quotation marks), and \eqref{fractionalK} is the precise, gauge-invariant description of the model. 
This is similar to the construction in \cite{Freed:2017rlk,Choi:2022jqy,Choi:2022rfe,Choi:2022fgx}.

\bibliographystyle{JHEP}
\bibliography{ref}

\end{document}